\documentclass[fleqn,usenatbib]{mnras}
\usepackage{newtxtext}
\usepackage[T1]{fontenc}
\usepackage{ae,aecompl}

\usepackage{graphicx}	

\usepackage{amssymb,amsmath}

\def\sqiglt{\hbox{\rlap{\lower.55ex \hbox {$\sim$}}\kern-.05em \raise.4ex \hbox{$<$}\,}}
\def\sqiggt{\hbox{\rlap{\lower.55ex \hbox {$\sim$}}\kern-.05em \raise.4ex \hbox{$>$}\,}}
\def\til{\ensuremath{\sim\,}}

\newcommand{\tim}[1]{\ensuremath{\times 10^{#1}}}
\def\deg{\ensuremath{^{\circ}}}

\def\cms{\ensuremath{$cm$^{-2}}}

\def\swift{\emph{Swift}}

\def\t0{\ensuremath{T_{0}}}

\def\P{\ensuremath{\mathcal{P}}}

\def\arcmin{\ensuremath{^\prime}}
\def\nu{\ensuremath{\nu}}

\def\dz{\ensuremath{\Delta z}}

\title[Swift observations during aLIGO O1]{\swift\ follow-up of gravitational wave triggers: results from the first aLIGO run 
and optimisation for the future}

\author[Evans et al.]{P.A. Evans$^1$\thanks{pae9@leicester.ac.uk}, J.A. Kennea$^2$, D.M. Palmer$^3$, M. Bilicki$^{4,5}$, J.P. Osborne$^1$, P.T. O'Brien$^1$, 
\and N.R. Tanvir$^1$, A.Y. Lien$^6$, S.D. Barthelmy$^6$, D. N. Burrows$^2$, S. Campana$^7$, S.B. Cenko$^{6,8}$
\and V. D'Elia$^{9,10}$, N. Gehrels$^6$, F. E. Marshall$^6$, K.L. Page$^1$, M.Perri$^{9,10}$, B. Sbarufatti$^{2,7}$, 
\and M.H. Siegel$^2$, G. Tagliaferri$^7$, E. Troja$^{6,11}$
\\
\\
$^1$Department of Physics and Astronomy, University of Leicester, Leicester, LE1 7RH, UK \\
$^2$Department of Astronomy and Astrophysics, Pennsylvania State University, 525 Davey Lab, University Park, PA 16802, USA \\
$^3$Los Alamos National Laboratory, B244, Los Alamos, NM, 87545, USA \\ 
$^4$Leiden Observatory, Leiden University, PO. Box 9513 Nl-2300 RA Leiden, The Netherlands \\
$^5$Janusz Gil Institute of Astronomy, University of Zielona G\'ora, ul. Lubuska 2, 65-265 Zielona G\'ora, Poland \\
$^6$NASA Goddard Space Flight Center, 8800 Greenbelt Road, Greenbelt, DMD 20771, USA \\
$^7$INAF, Osservatorio Astronomico di Brera, via E. Bianchi 46, 23807 Merate, Italy \\
$^8$Joint Space-Science Institute, University of Maryland, College Park, MD 20742, USA \\
$^9$INAF-Osservatorio Astronomico di Roma, via Frascati 33, I-00040 Monte Porzio Catone (RM), Italy \\
$^{10}$ASI-Science Data Center, Via del Politecnico snc, I-00133 Rome, Italy \\ 
$^{11}$Department of Physics and Astronomy, University of Maryland, College Park, MD 20742-4111, USA \\ 
}

\date{Accepted -- Received --}

\pagerange{\pageref{firstpage}--\pageref{lastpage}}
\pubyear{2016}

\begin{document}

\maketitle

\label{firstpage}

\begin{abstract} 
During its first observing run, in late 2015, the advanced LIGO facility announced 3 gravitational
wave (GW) triggers to electromagnetic follow-up partners. Two of these have since been confirmed
as being of astrophysical origin: both are binary black hole mergers at \til500 Mpc; the other trigger
was later found not to be astrophysical. In this paper we report on the \swift\ follow up observations of the second
and third triggers, including details of 21 X-ray
sources detected; none of which can be associated with the GW event. We also consider the challenges that
the next GW observing run will bring as the sensitivity and hence typical distance of GW events will increase.
We discuss how to effectively use galaxy catalogues to prioritise areas for follow up, especially in the
presence of distance estimates from the GW data. We also consider two galaxy catalogues and suggest that 
the high completeness at larger distances of the 2MASS Photometric Redshift Catalogue (2MPZ) makes it very well suited to
optimise \swift\ follow-up observations.
\end{abstract}


\section{Introduction}
\label{sec:intro}

In the last quarter of 2015, the Advanced Laser Interferometer Gravitational-wave Observatory observatory (aLIGO; 
\citealt{Aasi15,Abbott16g}) performed its first observing run (`O1') searching for gravitational waves (GW).
Each potential GW event was assigned a false alarm rate (FAR) indicating the frequency
with which a noise event with a signal of the observed strength is expected to arise.
Partner electromagnetic (EM) facilities, including \swift\ \citep{GehrelsSwift}, were notified of GW signals with an FAR
of less than one per month \citep{Abbott16EM}. O1 yielded the detection of two GW events, which have been confidently identified
as binary black hole (BBH) mergers: GW150914 \citep{Abbott16} and GW151226 \citep{Abbott16h},
and there was a further trigger (G194575) from the online analysis which was later determined to be a noise event.
Another possible merger event was detected in offline analysis of the O1 data [LVT151012; \mbox{\citep{Abbott16e}}].
Details of that event were not provided to EM partners until 2016 April, so no \swift\ follow up was performed.
The full results of O1 were reported by \cite{Abbott16O1}. 

Whilst the direct detection of GW was a significant achievement
which marked the beginning of a new era of astronomy, in order to maximise the scientific potential
of such discoveries, complementary EM data are needed. The three events
reported so far are all believed to be stellar-mass BBH mergers, which were not expected to produce
significant EM emission \citep[e.g.][]{Kamble13}. However, \emph{Fermi-GBM} reported
a possible low-significance event 0.4 s after the GW trigger for GW150914, which may be associated with
the GW event \citep{Connaughton16}. In the days following
the announcement of this, many authors suggested that EM emission from stellar-mass
BBH is possible given the correct binary parameters, or a charged black hole \citep[e.g.][]{Zhang16,Yamazaki16,Perna16,Loeb16},
although others suggested that a physical association between the GW and GBM events was unlikely \citep{Lyutikov16}.
Further, \emph{INTEGRAL} reported no detection \citep{Savchenko16} and suggest that this casts
doubt over whether the object detected by GBM was astrophysical in origin.
This issue will likely only be resolved by future GW detections of BBH with both contemporaneous and
follow-up EM observations.

Regardless of whether BBH mergers give rise to EM emission,
aLIGO is also expected to detect GW from the coalescence of binary neutron star  systems
or neutron star-black hole systems. These are both expected to produce multi-wavelength
EM radiation, for example in the form of a short gamma-ray burst (sGRB, e.g.\ \citealt{Berger14}) if the binary
is viewed close to face-on, or a kilonova \citep{Li98} regardless of the jet orientation;
see e.g.\ \cite{Nakar11,Metzger12,Zhang13} for a discussion of possible EM counterparts to such events.

In an earlier paper (\citealt{Evans16b}; hereafter `paper I') we presented the \swift\ observations
of GW150914. In this work we present the results of the \swift\ observations of the other two triggers reported
to the EM teams during O1, and consider how the \swift\ follow-up strategy may best evolve for the second run (O2) expected
in the second half of 2016. 

Throughout this paper, all errors are given at the 1-$\sigma$ level, unless stated otherwise.

\section{\swift\ observations}
\label{sec:response}

The \swift\ satellite \citep{GehrelsSwift} contains three complementary instruments. The
Burst Alert Telescope (BAT; \citealt{BarthelmyBAT}) is a 15--350 keV coded-mask instrument with a
field of view \til2 sr. Its primary role is to trigger on new transient events such as
GRBs. The other two instruments are narrow-field instruments, used for example to follow up
GRBs detected by BAT. The X-ray telescope (XRT; \citealt{BurrowsXRT}) is a 0.3--10 keV focusing instrument
with a peak effective area of 110 cm$^2$ at 1.5 keV and a roughly circular
field of view with radius 12.3\arcmin. The ultra-violet/optical telescope (UVOT; \citealt{RomingUVOT})
has 6 optical filters covering 1600--6240 \AA\ and a \emph{white} filter covering 1600--8000 \AA), with a peak 
effective area of 50 cm$^2$ in the $u$ band. The field of view is square, \til17\arcmin\ to a side.

The ideal scenario for \swift\ to observe a GW event would be for BAT to detect EM  emission (e.g.\ an sGRB) independently
of the GW trigger on the same event. \swift\ would then automatically slew and gather prompt
XRT and UVOT data. An sGRB is only seen if the coalescing binary is inclined such that the jet is 
oriented towards Earth; the opening angles of sGRB jets are not well known, however the observational
limits are in the range \til5--25\deg \citep{Burrows06,Grupe06,Fong15,Zhang15,Troja16}, therefore from purely 
geometrical constraints, we expect only a minority of binary neutron star/neutron star-black hole mergers detected in GW to be
accompanied by an sGRB (e.g.\ for a jet angle of 10\deg, only 1.5\%\ will be viewed on-axis), whereas the GW signal is only 
modestly affeted by binary inclination.  
Some authors (e.g.\ \citealt{Troja10,Tsang13}) have
suggested that the neutron star crust can be disrupted prior to the merger and that this 
could give rise to an isotropic precursor, i.e.\ BAT could in principle detect such emission
from an off-axis GRB. However, these could well be too faint to trigger \swift-BAT.
Also, while an excellent GRB-detection machine, \swift-BAT can only 
observe \til1/6 of the sky at any given time. The combination of these factors means that, while
a simultaneous ALIGO-BAT detection would be scientifically optimal, it is not a particularly likely
occurrance.

In addition, \swift\ can respond to the GW trigger, and observe a portion of the GW error region
(which typically hundreds of square degrees in size) rapidly with its narrow-field telescopes. \cite{Evans16} discussed
optimal ways to do this, focusing primarily on the XRT, since it has a larger
field of view than the UVOT, and the expected rate of unrelated transient events
in the X-ray range, while not well constrained, is expected to be lower than in the optical
bands (see \citealt{Kanner13,Evans16} for a discussion of X-ray transient rates). Their suggested
approach was to modify the GW error region by means of a galaxy catalogue (they used the Gravitational Wave 
Galaxy Catalogue (GWGC): \citealt{White11}), weighting each pixel in the GW skymap by the luminosity 
of the catalogued galaxies in that pixel,  and then to observe in a succession of short observations, in decreasing
order of probability in this combined map. A more detailed Bayesian approach to this was discussed
by \cite{Fan14}. As we reported in paper I, the ability to observe a large number of fields with short exposures required operational changes
for \swift\ which were not completed in time for O1, therefore we  were only 
able to observe a relatively small part of the GW error regions for the triggers during that run.
As described in paper I for GW150914, we combined the GW error region for each trigger in O1 with GWGC, weighting the galaxies
according to their $B$-band luminosities, and selecting XRT fields based on the resultant
probability map.

The data analysis approach was described in paper I so
here we offer only a pr\'ecis. For XRT, the source detection system was based on that of \cite{Evans14}, slightly modified to support
shorter exposures. Every source detected was automatically assigned a \emph{Rank} of 1--4 describing
how likely it was to be the counterpart to the GW event, with 1 being the most likely. This was based on
whether the source was previously catalogued, its flux compared to previous detection or upper limits
and its proximity to a known galaxy (full definitions of the ranks are in paper I).
The detection system also produces warning flags for sources which
it believes may be spurious due to effects, such as diffuse X-ray emission (the detection system 
is designed for point sources), or instrumental artefacts, such as stray light or optical
loading (see \citealt{Evans14} section~3.4 and fig.~5). For each source detected, a GCN `Counterpart'
notice was automatically produced as soon as the source was detected; this contained standard details 
(position, time of detection, flux) and also the rank and any warning flags\footnote{For most of O1 these extra
fields were only included in the email form of the GCN notice. Towards the end of O1 these
were also added to the binary-format notice.}. All sources were checked by humans, and any which 
were spurious were removed, and the verified sources reported in LVC/GCN circulars
(\citealt{EvansLVCCirc1,EvansLVC_18346,EvansLVC_18446,EvansLVC_18732,EvansLVC_18733,EvansLVC_18748,EvansLVC_18834,EvansLVC_18870}).

UVOT data were analysed using standard {\sc heasoft} tools, and an automated pipeline was used to search
for transients. Visual screening was applied to UVOT images, using the Digitized Sky Survey as a
comparison. Although no rank 1 or 2 X-ray sources were found during O1, the UVOT data around
any such sources would also have been closely inspected by eye.

\subsection{GW150914}
\label{sec:obs150914}

GW150914 was detected at the end of the aLIGO engineering run immediately prior to O1, and
was the first ever direct detection of GW. The \swift\ results for this event were reported
in full in paper I. Recently (2016 April) we reobserved the GW error region of GW150914, as
the final step to commission the ability to perform large-scale rapid tiling with \swift. \swift\ observed 426 fields during the UT day 2016 April 21
with 60 s of exposure per field\footnote{The GW error region is not observable for the entire \swift\ orbit,
which is why the total exposure was $426\times60$ s $\ll 1$ day.}, covering a total of 53 sq deg. Only one X-ray source was
detected in these observations, the known X-ray emitter 1RXS J082709.9$-$650447, which was detected with a flux 
consistent with that from the \emph{Rosat} observations \citep{Voges99}. The scientific `result' from these observations
is that, as expected, the only background X-ray source found was a known (rank 4) object; therefore we are optimistic
that a transient should be easy to distinguish. However, these observations also demonstrate that \swift\ is now
capable of performing large-scale tiling in response to a GW trigger.

\subsection{Trigger G194575}
\label{sec:obs151022}

\begin{figure}
\begin{center}
\includegraphics[width=8.1cm,angle=0]{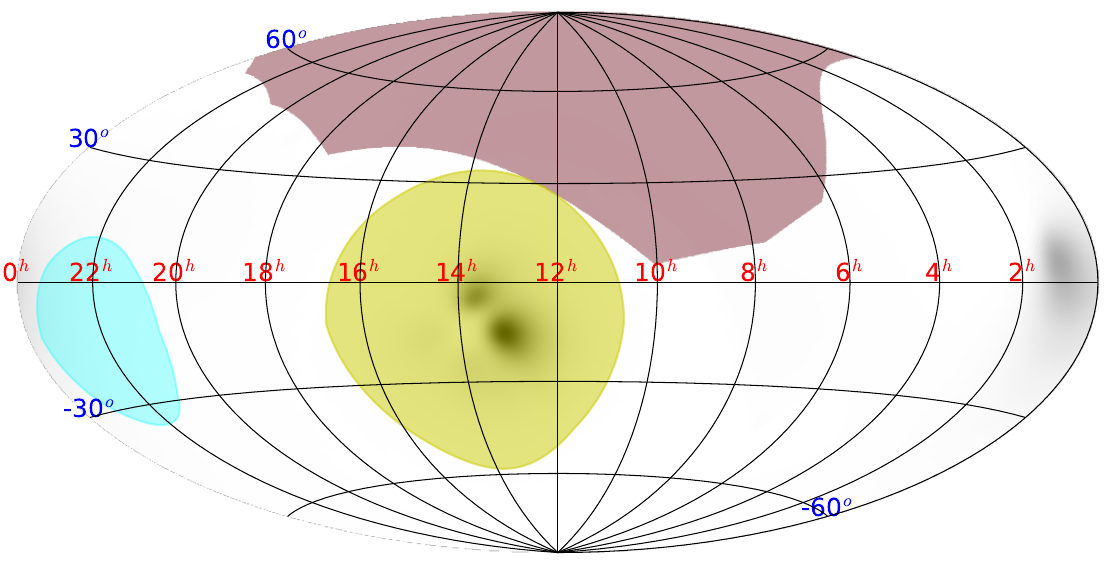}
\includegraphics[width=8.1cm,angle=0]{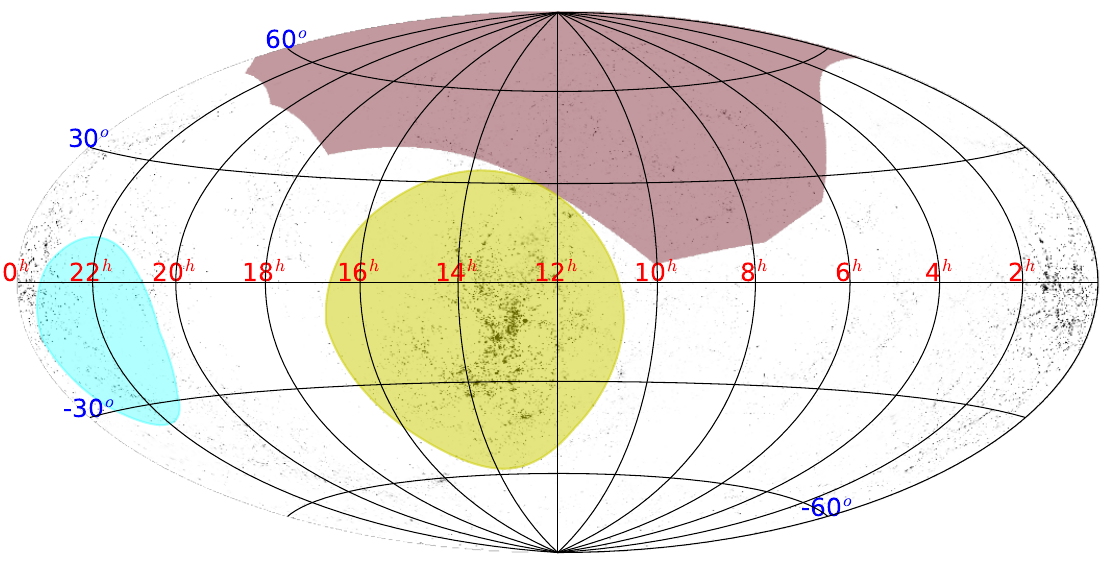}
\end{center}
\caption{The `BAYESTAR' GW localisation map for trigger G194575, produced by the LVC team on 2015 October 22 (top), combined with our
luminosity-weighted GWGC map (bottom). Coordinates are equatorial, J2000.
The yellow and cyan circles indicate the regions towards which XRT and UVOT cannot
point due to proximity to the Sun and Moon respectively. The large maroon area is the BAT partially coded field of view
at the time of the GW event.
For this event, unfortunately the majority of the error region was unobservable due to its proximity to the Sun.}
\label{fig:151022_skymap}
\end{figure}

\begin{table*}
\begin{center}
\caption{\swift\ observations of the error region of LVC trigger G194575}
\label{tab:obs_151022}
\begin{tabular}{cccccc}
\hline
Pointing direction & Start time$^a$ & Exposure & Source & XRT limit & UVOT magnitude\\
(J2000) & (UTC) & (s) & & 0.3--10 keV \\
& & & & erg \cms\ s$^{-1}$ \\
\hline
$00^h 11^m 27.60^s, -06\deg 25^\prime 38.3^{\prime\prime}$  &  Oct 27 01:17:46  &   1985 & LSQ15bjb & 1.4\tim{-13} & $u$=16.7 \\
$00^h 58^m 13.27^s, -03\deg 39^\prime 50.4^{\prime\prime}$  &  Nov 06 23:22:15  &   9948$^b$ & iPTF15dld  & 4.9\tim{-14} & N/A$^c$\\
\hline
\end{tabular}
\end{center}
\begin{flushleft}
$^a$ All observations were in 2015.
\newline
$^b$ The observation of iPTF15dld was not a continuous exposure due to \swift's low-Earth orbit. The 10 ks of data
were obtained between the Nov 6 at 23:22:15 and Nov 07 at 10:16:44 UT.
\newline $^c$ The source could not be deconvolved from the host galaxy in the UVOT data, so no magnitude was derived.
\end{flushleft}
\end{table*} 

The aLIGO `compact binary coalescence' (CBC) pipeline, which uses a template library
of expected GW waveforms from merging compact binaries, triggered on 2015 October 22 at 
13:33:19.942 UT. The detected signal had an FAR of 
9.65\tim{-8} Hz, equivalent to one per four months \citep{Singer15b}.
Unfortunately, most of the higher probability areas of the error region were 
too close to the Sun for observations with \swift\ (Fig.~\ref{fig:151022_skymap}),
therefore only the low probability regions were observable. Additionally, offline analysis
of the GW signal reduced the FAR to 8.19\tim{-6} Hz (one per 1.41 
days), and it was therefore determined not to be a real GW event \citep{LSC_18626}.
Although this trigger is therefore not of astrophysical significance, one point
of procedural interest for future triggers is worth noting. 
Before the significance of the trigger had been downgraded, two sources identified
by ground-based observatories were reported as being of potential interest: 
LSQ15bjb \citep{Rabinowitz15} and iPTF15dld \citep{Singer15c}, which 
were detected by the La~Silla QUEST and iPTF ground-based facilities 
respectively. \swift\ observed both of these sources and, finding no X-ray counterpart, we
reported upper limits \citep{EvansLVC_18489,EvansLVC_18569}. LSQ15bjb was originally reported as an
uncatalogued and rapidly brightening optical source \citep{Rabinowitz15}, which was subsequently 
classified as a type Ia supernova \citep{Piranomonte15}. iPTF15dld was one of
several optical transients reported by \citep{Singer15c} that were consistent in position
with a known galaxy at $z<0.1$; it later transpired that this source had been detected by
by La~Silla QUEST 19 days before the GW event \citep{Rabinowitz15b}.

Details of the \swift\ observations and results for these two sources are given in Table~\ref{tab:obs_151022}.
This demonstrates the ability to be flexible when performing \swift\ GW follow up, and
perform targetted observation of point sources detected by other facilities, as well 
the blind searches.



\subsection{GW 151226}
\label{sec:obs151226}

\begin{figure}
\begin{center}
\includegraphics[width=8.1cm,angle=0]{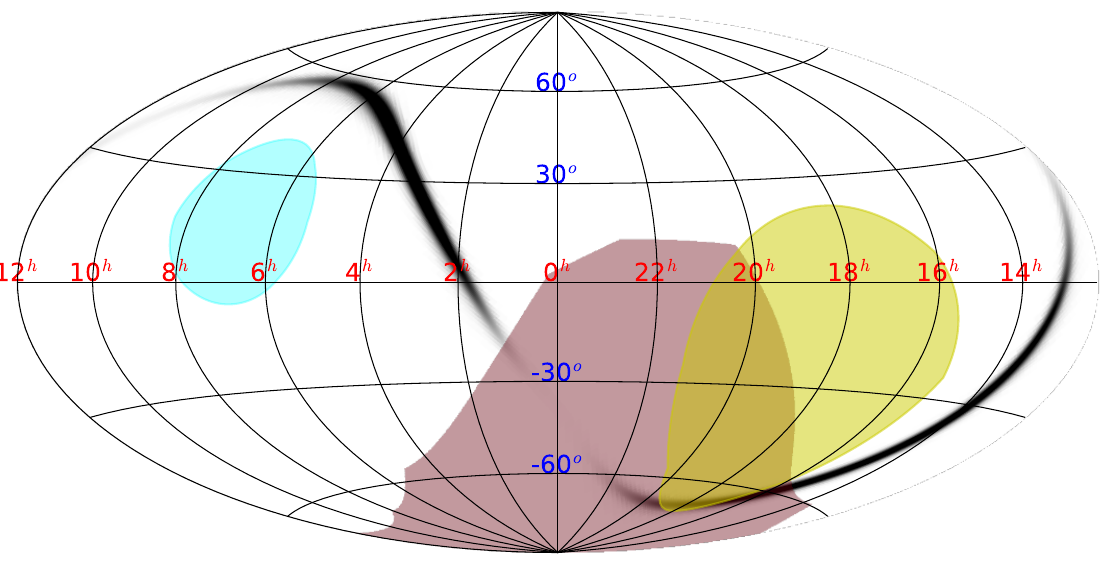}
\includegraphics[width=8.1cm,angle=0]{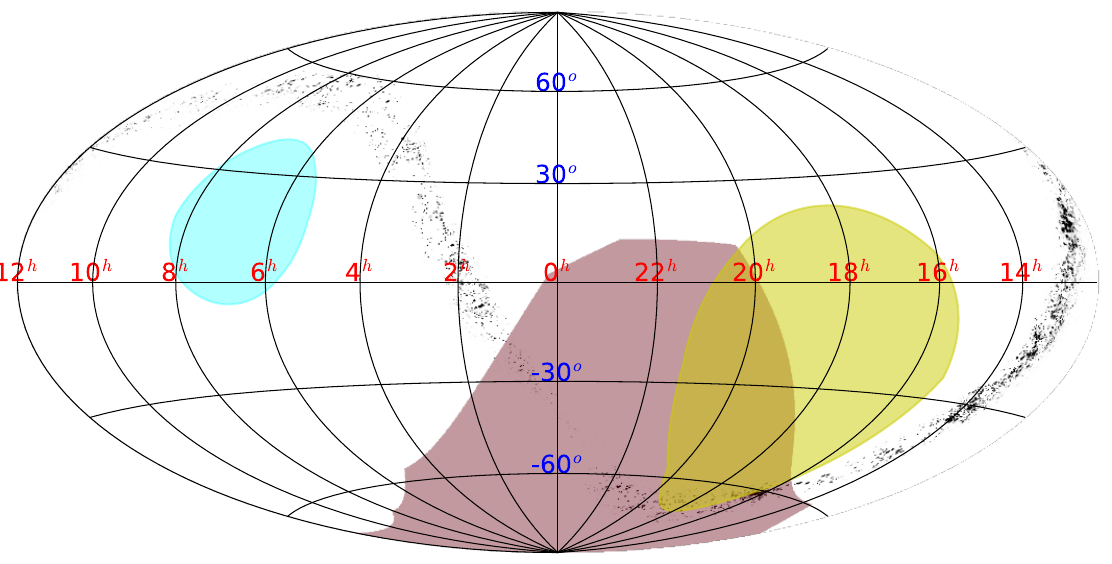}
\includegraphics[width=8.1cm,angle=0]{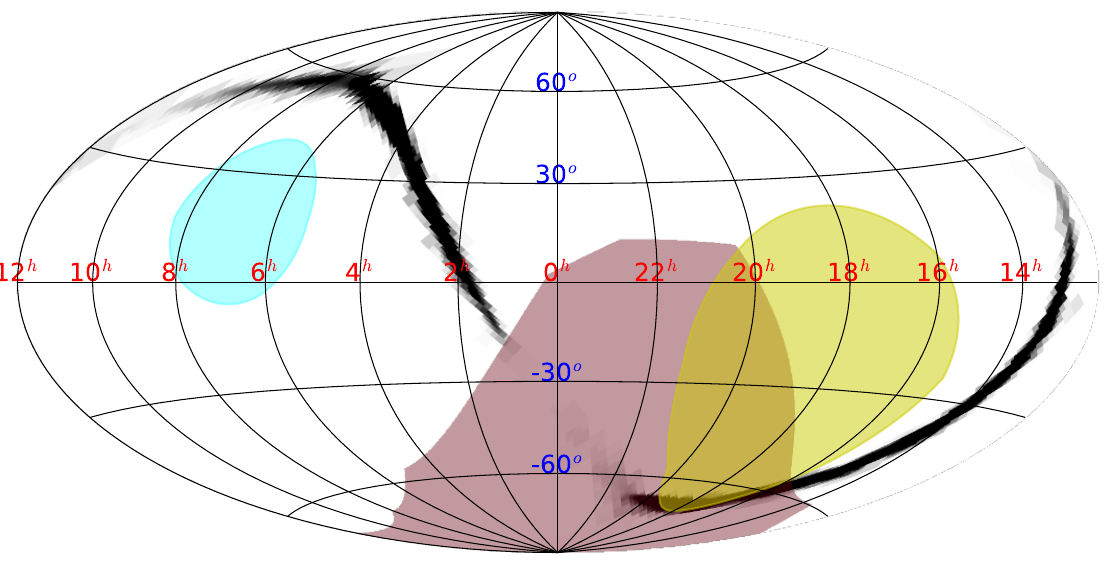}
\end{center}
\caption{The `BAYESTAR' GW localisation map of GW151226, produced by the LVC team on 2015 December 26 (top), combined with our
luminosity-weighted GWGC map (middle). The bottom panel is the refined `LALInference` map. The yellow and cyan circles are as in Fig.~\ref{fig:151022_skymap}.
These images are centred on RA=0, unlike Fig.~\ref{fig:151022_skymap}, so that the regions are more visible.}
\label{fig:151226_skymap}
\end{figure}

\begin{table}
\begin{center}
\caption{\swift\ observations of the error region of GW 151226.}
\label{tab:obs_151226}
\begin{tabular}{ccc}
\hline
Pointing direction & Start time$^a$ & Exposure \\
(J2000) & (UTC) & (s) \\
\hline
$09^h 43^m 50.88^s, +59\deg 48^\prime 02.9^{\prime\prime}$  &  Dec 27 at 18:37:03  &   1384  \\
$13^h 30^m 7.20^s, -21\deg 13^\prime 01.2^{\prime\prime}$  &  Dec 27 at 20:19:11  &   333  \\
$13^h 31^m 33.84^s, -21\deg 13^\prime 01.2^{\prime\prime}$  &  Dec 27 at 20:21:31  &   325  \\
$13^h 30^m 50.64^s, -21\deg 30^\prime 32.4^{\prime\prime}$  &  Dec 27 at 20:23:53  &   288  \\
$13^h 29^m 23.76^s, -21\deg 30^\prime 32.4^{\prime\prime}$  &  Dec 27 at 20:26:14  &   318  \\
$13^h 28^m 40.56^s, -21\deg 13^\prime 01.2^{\prime\prime}$  &  Dec 27 at 20:28:34  &   305  \\
$13^h 29^m 23.76^s, -20\deg 55^\prime 30.0^{\prime\prime}$  &  Dec 27 at 20:30:53  &   313  \\
$13^h 30^m 50.64^s, -20\deg 55^\prime 30.0^{\prime\prime}$  &  Dec 27 at 20:33:13  &   340  \\
$13^h 32^m 17.28^s, -20\deg 55^\prime 30.0^{\prime\prime}$  &  Dec 27 at 20:35:30  &   213  \\
$13^h 33^m 0.48^s, -21\deg 13^\prime 01.2^{\prime\prime}$  &  Dec 27 at 20:37:49  &   285  \\
$13^h 32^m 17.28^s, -21\deg 30^\prime 32.4^{\prime\prime}$  &  Dec 27 at 20:40:07  &   310  \\
$13^h 31^m 33.84^s, -21\deg 48^\prime 00.0^{\prime\prime}$  &  Dec 27 at 20:42:25  &   325  \\
$13^h 30^m 7.20^s, -21\deg 48^\prime 00.0^{\prime\prime}$  &  Dec 27 at 20:44:42  &   378  \\
$13^h 28^m 40.56^s, -21\deg 48^\prime 00.0^{\prime\prime}$  &  Dec 27 at 20:46:59  &   318  \\
$13^h 27^m 57.12^s, -21\deg 30^\prime 32.4^{\prime\prime}$  &  Dec 27 at 20:49:11  &   320  \\
$13^h 27^m 13.68^s, -21\deg 13^\prime 01.2^{\prime\prime}$  &  Dec 27 at 20:51:23  &   323  \\
$13^h 27^m 57.12^s, -20\deg 55^\prime 30.0^{\prime\prime}$  &  Dec 27 at 20:53:32  &   320  \\
$13^h 28^m 40.56^s, -20\deg 38^\prime 02.4^{\prime\prime}$  &  Dec 27 at 20:55:39  &   308  \\
$13^h 30^m 7.20^s, -20\deg 38^\prime 02.4^{\prime\prime}$  &  Dec 27 at 20:57:44  &   323  \\
$13^h 31^m 33.84^s, -20\deg 38^\prime 02.4^{\prime\prime}$  &  Dec 27 at 20:59:47  &   290  \\
$13^h 49^m 32.88^s, -30\deg 29^\prime 16.8^{\prime\prime}$  &  Dec 28 at 01:14:23  &   308  \\
$13^h 51^m 6.72^s, -30\deg 29^\prime 16.8^{\prime\prime}$  &  Dec 28 at 01:16:22  &   193  \\
$13^h 50^m 19.92^s, -30\deg 46^\prime 48.0^{\prime\prime}$  &  Dec 28 at 01:18:21  &   157  \\
$13^h 48^m 46.08^s, -30\deg 46^\prime 48.0^{\prime\prime}$  &  Dec 28 at 01:20:21  &   295  \\
$13^h 47^m 59.04^s, -30\deg 29^\prime 16.8^{\prime\prime}$  &  Dec 28 at 01:22:19  &   310  \\
$13^h 48^m 46.08^s, -30\deg 11^\prime 49.2^{\prime\prime}$  &  Dec 28 at 01:24:16  &   438  \\
$13^h 50^m 19.92^s, -30\deg 11^\prime 49.2^{\prime\prime}$  &  Dec 28 at 01:26:12  &   300  \\
$13^h 51^m 53.76^s, -30\deg 11^\prime 49.2^{\prime\prime}$  &  Dec 28 at 01:28:08  &   310  \\
$13^h 52^m 40.56^s, -30\deg 29^\prime 16.8^{\prime\prime}$  &  Dec 28 at 01:30:04  &   423  \\
$13^h 51^m 53.76^s, -30\deg 46^\prime 48.0^{\prime\prime}$  &  Dec 28 at 01:32:00  &   165  \\
$13^h 51^m 6.72^s, -31\deg 04^\prime 19.2^{\prime\prime}$  &  Dec 28 at 01:33:55  &   290  \\
$13^h 49^m 32.88^s, -31\deg 04^\prime 19.2^{\prime\prime}$  &  Dec 28 at 01:35:49  &   305  \\
$13^h 47^m 59.04^s, -31\deg 04^\prime 19.2^{\prime\prime}$  &  Dec 28 at 01:37:42  &   313  \\
$13^h 47^m 12.24^s, -30\deg 46^\prime 48.0^{\prime\prime}$  &  Dec 28 at 01:39:33  &   418  \\
$13^h 46^m 25.44^s, -30\deg 29^\prime 16.8^{\prime\prime}$  &  Dec 28 at 01:41:22  &   245  \\
$13^h 47^m 12.24^s, -30\deg 11^\prime 49.2^{\prime\prime}$  &  Dec 28 at 01:43:08  &   368  \\
$13^h 47^m 59.04^s, -29\deg 54^\prime 18.0^{\prime\prime}$  &  Dec 28 at 01:44:53  &   303  \\
$13^h 49^m 32.88^s, -29\deg 54^\prime 18.0^{\prime\prime}$  &  Dec 28 at 01:46:37  &   165  \\
$13^h 51^m 6.72^s, -29\deg 54^\prime 18.0^{\prime\prime}$  &  Dec 28 at 01:48:19  &   127  \\
$14^h 03^m 9.12^s, -34\deg 09^\prime 25.2^{\prime\prime}$  &  Dec 28 at 09:14:07  &   80  \\
$14^h 04^m 47.04^s, -34\deg 09^\prime 25.2^{\prime\prime}$  &  Dec 28 at 09:16:09  &   215  \\
$14^h 03^m 58.08^s, -34\deg 26^\prime 52.8^{\prime\prime}$  &  Dec 28 at 09:18:05  &   315  \\
$14^h 02^m 20.40^s, -34\deg 26^\prime 52.8^{\prime\prime}$  &  Dec 28 at 09:20:03  &   218  \\
$14^h 01^m 31.68^s, -34\deg 09^\prime 25.2^{\prime\prime}$  &  Dec 28 at 09:21:59  &   87  \\
$14^h 02^m 20.40^s, -33\deg 51^\prime 54.0^{\prime\prime}$  &  Dec 28 at 09:23:54  &   115  \\
$14^h 03^m 58.08^s, -33\deg 51^\prime 54.0^{\prime\prime}$  &  Dec 28 at 09:25:49  &   90  \\
$14^h 05^m 35.76^s, -33\deg 51^\prime 54.0^{\prime\prime}$  &  Dec 28 at 09:27:42  &   105  \\
$14^h 06^m 24.48^s, -34\deg 09^\prime 25.2^{\prime\prime}$  &  Dec 28 at 09:29:37  &   100  \\
$14^h 05^m 35.76^s, -34\deg 26^\prime 52.8^{\prime\prime}$  &  Dec 28 at 09:31:31  &   92  \\
$14^h 04^m 47.04^s, -34\deg 44^\prime 24.0^{\prime\prime}$  &  Dec 28 at 09:33:25  &   107  \\
$14^h 01^m 31.68^s, -34\deg 44^\prime 24.0^{\prime\prime}$  &  Dec 28 at 09:37:08  &   285  \\
$14^h 00^m 42.72^s, -34\deg 26^\prime 52.8^{\prime\prime}$  &  Dec 28 at 09:38:57  &   135  \\
$13^h 59^m 54.00^s, -34\deg 09^\prime 25.2^{\prime\prime}$  &  Dec 28 at 09:40:44  &   295  \\
\hline
\end{tabular}
\end{center}
\begin{flushleft}
$^a$ Observations were 2015 December or 2016 January.
\end{flushleft}
\end{table}

\begin{table}
\begin{center}
\contcaption{}
\begin{tabular}{ccc}
\hline
Pointing direction & Start time$^a$ & Exposure \\
(J2000) & (UTC) & (s) \\
\hline
$14^h 00^m 42.72^s, -33\deg 51^\prime 54.0^{\prime\prime}$  &  Dec 28 at 09:42:31  &   310  \\
$14^h 01^m 31.68^s, -33\deg 34^\prime 22.8^{\prime\prime}$  &  Dec 28 at 09:44:13  &   195  \\
$14^h 03^m 9.12^s, -33\deg 34^\prime 22.8^{\prime\prime}$  &  Dec 28 at 09:45:56  &   401  \\
$14^h 04^m 47.04^s, -33\deg 34^\prime 22.8^{\prime\prime}$  &  Dec 28 at 09:47:37  &   278  \\
$12^h 31^m 7.44^s, +12\deg 18^\prime 50.4^{\prime\prime}$  &  Dec 28 at 15:24:31  &   443  \\
$12^h 32^m 30.24^s, +12\deg 18^\prime 50.4^{\prime\prime}$  &  Dec 28 at 15:26:54  &   406  \\
$12^h 31^m 48.96^s, +12\deg 01^\prime 22.8^{\prime\prime}$  &  Dec 28 at 15:29:21  &   438  \\
$12^h 30^m 26.16^s, +12\deg 01^\prime 22.8^{\prime\prime}$  &  Dec 28 at 15:31:47  &   431  \\
$12^h 29^m 44.88^s, +12\deg 18^\prime 50.4^{\prime\prime}$  &  Dec 28 at 15:34:11  &   423  \\
$12^h 30^m 26.16^s, +12\deg 36^\prime 21.6^{\prime\prime}$  &  Dec 28 at 15:36:36  &   386  \\
$12^h 31^m 48.96^s, +12\deg 36^\prime 21.6^{\prime\prime}$  &  Dec 28 at 15:38:53  &   418  \\
$12^h 33^m 11.52^s, +12\deg 36^\prime 21.6^{\prime\prime}$  &  Dec 28 at 15:41:09  &   416  \\
$12^h 33^m 53.04^s, +12\deg 18^\prime 50.4^{\prime\prime}$  &  Dec 28 at 15:43:22  &   423  \\
$12^h 33^m 11.52^s, +12\deg 01^\prime 22.8^{\prime\prime}$  &  Dec 28 at 15:45:34  &   413  \\
$12^h 32^m 30.24^s, +11\deg 43^\prime 51.6^{\prime\prime}$  &  Dec 28 at 15:47:47  &   421  \\
$12^h 31^m 7.44^s, +11\deg 43^\prime 51.6^{\prime\prime}$  &  Dec 28 at 15:49:58  &   433  \\
$12^h 29^m 44.88^s, +11\deg 43^\prime 51.6^{\prime\prime}$  &  Dec 28 at 15:52:08  &   411  \\
$12^h 29^m 3.12^s, +12\deg 01^\prime 14.7^{\prime\prime}$  &  Dec 28 at 15:54:15  &   423  \\
$12^h 28^m 22.08^s, +12\deg 18^\prime 50.4^{\prime\prime}$  &  Dec 28 at 15:56:21  &   418  \\
$12^h 29^m 3.36^s, +12\deg 36^\prime 21.6^{\prime\prime}$  &  Dec 28 at 15:58:23  &   428  \\
$12^h 29^m 44.88^s, +12\deg 53^\prime 52.8^{\prime\prime}$  &  Dec 28 at 16:00:24  &   335  \\
$12^h 31^m 7.44^s, +12\deg 53^\prime 52.8^{\prime\prime}$  &  Dec 28 at 16:02:46  &   263  \\
$12^h 32^m 30.24^s, +12\deg 53^\prime 52.8^{\prime\prime}$  &  Dec 28 at 16:05:01  &   328  \\
$02^h 59^m 41.20^s, +25\deg 14^\prime 12.2^{\prime\prime}$  &  Jan 05 at 17:43:10  &   3763$^b$ \\
$02^h 32^m 59.75^s, +18\deg 38^\prime 07.0^{\prime\prime}$  &  Jan 07 at 15:52:50  &   17182$^c$  \\
\end{tabular}
\end{center}
\begin{flushleft}
$^a$ Observations were 2015 December or 2016 January.
\newline
$^b$ Observations of PS15dqa. These observations were not continuous but occurred in two `snapshots' on consecutive \swift\ orbits.
\newline
$^c$ Observations of PS15dpn. This was observed every few days for two weeks, the last observation occurring on 2016 January 25.
\end{flushleft}
\end{table}

\begin{table*}
\begin{center}
\caption{XRT sources detected in observations of GW151226}
\label{tab:res_151226}
\begin{tabular}{cccclc}
\hline
Position & Error$^a$ & Flux (erg \cms\ s$^{-1}$) & Rank & Catalogued match  & Separation\\
(J2000) & (arcsec) & 0.3--10 keV &  & & (arcsec)\\
\hline
$13^h 30^m 13.26^s, -20\deg 54^\prime 16.4^{\prime\prime}$  & 5.3 & $(4.1\pm2.2)\tim{13}$  & 3 & \\
$12^h 30^m 47.32^s, +12\deg 20^\prime 20.3^{\prime\prime}$  & 6.6 & $(6.3\pm2.3)\tim{12}$  & 3 & \\
$12^h 32^m 7.03^s, +11\deg 51^\prime 21.6^{\prime\prime}$  & 6.6 & $(6.7\pm2.6)\tim{13}$  & 3 & \\
$13^h 30^m 16.30^s, -20\deg 55^\prime 26.1^{\prime\prime}$  & 6.0 & $(4.6\pm2.8)\tim{13}$  & 3 & \\
$12^h 31^m 29.59^s, +11\deg 52^\prime 37.8^{\prime\prime}$  & 6.7 & $(6.0\pm2.7)\tim{13}$  & 3 & \\
$12^h 31^m 42.62^s, +12\deg 19^\prime 45.3^{\prime\prime}$  & 13.1 & $(7.7\pm4.4)\tim{13}$& 3 & \\
$13^h 29^m 25.00^s, -21\deg 13^\prime 37.2^{\prime\prime}$  & 5.7 & $(1.1\pm0.4)\tim{12}$  & 3 & \\
$02^h 59^m 42.18^s, +25\deg 12^\prime 46.5^{\prime\prime}$  & 5.8 & $(6.3\pm3.4)\tim{14}$  & 3 & \\
$12^h 30^m 59.30^s, +12\deg 11^\prime 33.9^{\prime\prime}$  & 5.6 & $(8.3\pm4.6)\tim{13}$  & 4 & 3XMM J123059.4+121131 &  2.7  \\
$13^h 49^m 19.27^s, -30\deg 18^\prime 35.4^{\prime\prime}$  & 4.1 & $(5.3\pm1.2)\tim{11}$ & 4 & 1SXPS J134919.2-301834 &  6.5  \\
$13^h 48^m 44.40^s, -30\deg 29^\prime 46.5^{\prime\prime}$  & 5.9 & $(2.2\pm0.5)\tim{12}$  & 4 & XMMSL1 J134844.6-302948 &  2.7  \\
$13^h 30^m 7.66^s, -20\deg 56^\prime 11.1^{\prime\prime}$  & 6.0 & $(8.5\pm3.1)\tim{13}$  & 4 & 1SXPS J133007.7-205619 &  8.4  \\
 & & & & [RKV2003] QSO J1330-2056 abs 0.84992 & 5.3$^b$  \\
$12^h 30^m 42.99^s, +12\deg 23^\prime 17.1^{\prime\prime}$  & 5.5 & $(1.8\pm0.5)\tim{10}$ & 4 & 2RXP J123044.7+122331 &  28.9  \\
 & & & & [SFH81] 1157 & 7.9$^b$  \\
$13^h 30^m 7.02^s, -21\deg 41^\prime 59.8^{\prime\prime}$  & 5.1 & $(2.7\pm0.6)\tim{12}$  & 4 & 1RXS J133006.8-214156 &  3.9  \\
 & & & & [RKV2003] QSO J1330-2142 abs 0.3014 & 2.9$^b$  \\
$13^h 49^m 4.00^s, -30\deg 17^\prime 46.3^{\prime\prime}$  & 7.6 & $(2.0\pm0.5)\tim{12}$  & 4 & XMMSL1 J134904.4-301745 &  3.6  \\
 & & & & [RP98d] P6 & 6.3$^b$  \\
$12^h 31^m 12.74^s, +12\deg 03^\prime 17.1^{\prime\prime}$  & 7.3 & $(3.0\pm1.1)\tim{13}$  & 4 & 3XMM J123113.1+120307 &  6.6  \\
 & & & & 2MASX J12311311+1203075 &  6.6$^b$  \\

\hline
\end{tabular}
\end{center}
\begin{flushleft}
$^a$ 90\%\ confidence
\newline
$^b$ From SIMBAD
\end{flushleft}
\end{table*} 

\begin{figure}
\begin{center}
\includegraphics[width=8.1cm,angle=0]{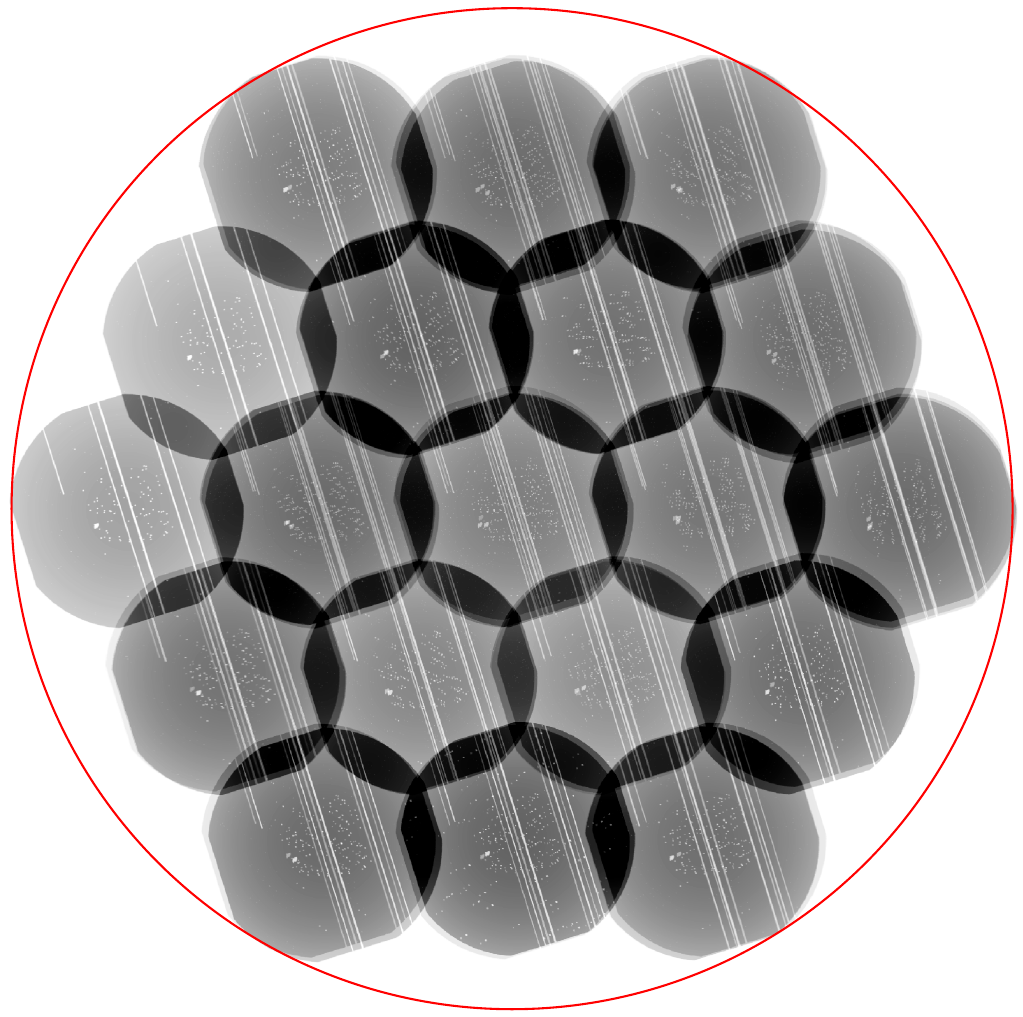}
\includegraphics[width=8.1cm,angle=0]{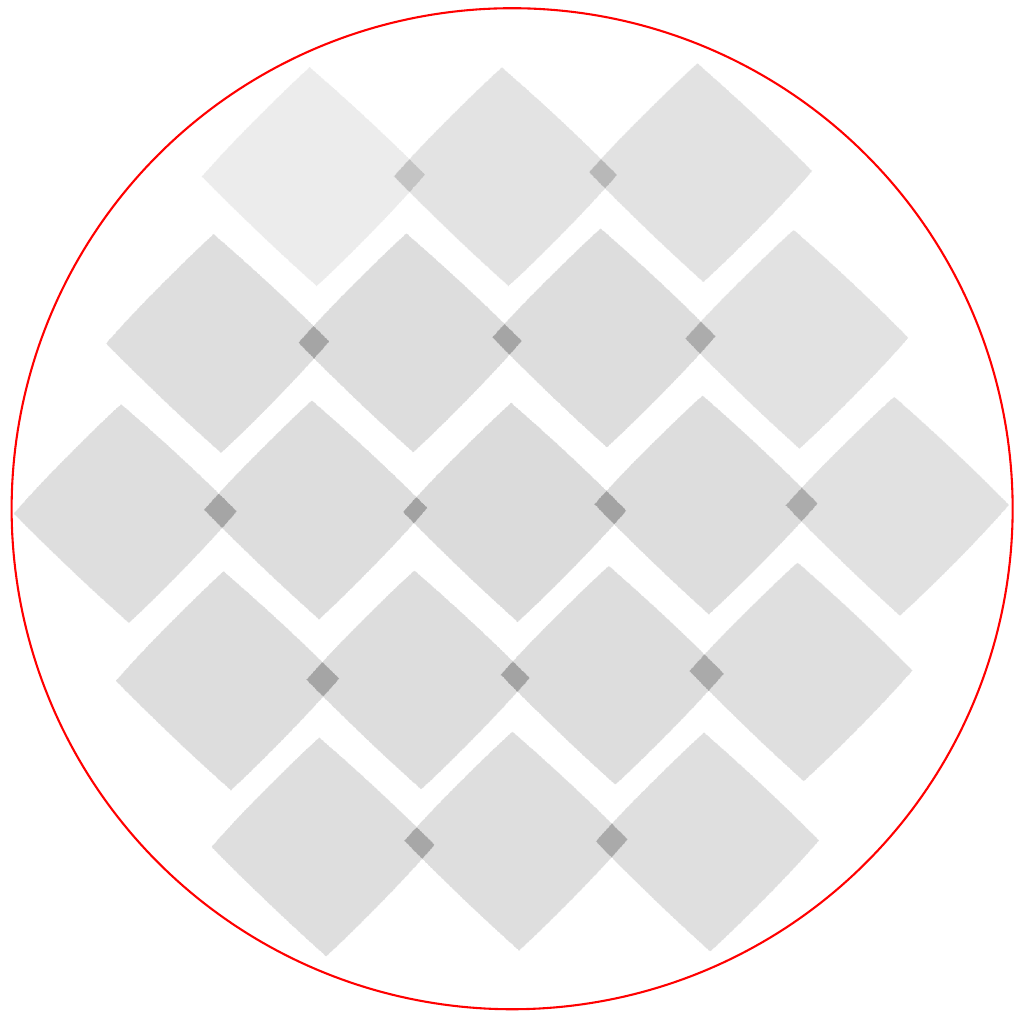}
\end{center}
\caption{An example XRT (top) and UVOT (bottom) exposure map from a 19-point tile used in the follow-up observations of GW151226.
The circle is shown for scale and has radius 0.88\deg.}
\label{fig:19point}
\end{figure}

The CBC pipeline triggered on 2015 December 26 at 03:38:53.648 UT, with a signal
with FAR lower than one per month \mbox{\citep{LSC_18851}}; this was later refined to an FAR
lower than one per hundred years \mbox{\citep{LSC_18853}}. The GW waveform indicated that this was a high-mass event,
most likely a BH-BH merger \mbox{\citep{LSC_18851}}. As with both previous triggers,
a portion of the error region was too close to the Sun to observe with \swift\ (Fig.~\ref{fig:151226_skymap}).
The trigger was announced to the follow-up community on 2015 December 27 at 16:28 UT, and \swift\ observations
began at 18:35 UT on the same day. We followed the same procedure as for the earlier triggers, selecting
the most probable XRT fields after combining with the GWGC galaxy catalogue. However, after the first field had been observed, we modified this approach 
and instead of selecting single XRT fields, we selected regions covered by a set of 19 tiled pointings (Fig.~\ref{fig:19point}). We uploaded four
such observation sets, as detailed in Table~\ref{tab:obs_151226}. We also performed additional sets of observations, observing the locations
of PS15dqa and PS15dpn \citep{Chambers15}: optical sources highlighted as potentially interesting. PS15dpn was observed repeatedly for several days in order
to track the UV evolution of its light curve. This source is not believed to be related to the GW trigger, and the PS15dpn data are not presented in this work. 

The GW localisation of GW151226 is shown in Fig.~\ref{fig:151226_skymap}. At the time of the \swift\ observations,
only the `BAYESTAR' map (produced by the low-latency pipeline; \citealt{Singer16b}) was available (top two panels). A revised skymap
produced by the offline `LALInference' pipeline [\cite{Veitch15}; bottom panel of Fig.~\ref{fig:151226_skymap}] was made
available on 2016 January 16 \mbox{\citep{LSC_18889}}, well after our observations had been completed. The BAT field of view overlaps the GW localisations, covering
14\%\ of the probability in the `BAYESTAR' map and 15\%\ from the revised map [these probabilities are higher, \til29\%\, and 33\% 
after weighting by the GWGC, however since the distance to this merger is large ($440^{+180}_{-190}$ Mpc; \citealt{Abbott16h})
and GWGC only contains galaxies to 100 Mpc, the galaxy-weighted map is not appropriate\footnote{The distance estimate from the GW
data was not available at the time of the observations.}].  We created a 15--350 keV BAT light curve from $T_0-100$ to $T_0+100$ s ($T_0$ is the 
GW trigger time) with bins of 1.024 s. No signal is seen, at the 4-$\sigma$ level with an upper limit (also 4 $\sigma$) of 303.6 counts in a single bin. We
used 4 $\sigma$ as the limit rather than 3 $\sigma$ because, for a 1.024-s binned light curve we expect a 3-$\sigma$ noise fluctuation every \til6 minutes,
therefore the chance of a spurious signal in our data is high. At 4 $\sigma$ noise fluctuations are expected only every 4.4 hours.
To convert this to a flux limit we assumed a typical short GRB BAT spectrum: a power-law with a photon index of 1.32.
The counts-to-flux conversion depends on the angle of the source to the BAT boresight which, since the source position is poorly
contsrained, is not known. If the GW source was close to the BAT boresight (`fully coded' by the BAT mask) the upper limit is 4.3\tim{-8} erg \cms\ s$^{-1}$.
For a source which is half coded by the mask (45\deg\ off axis) the limit is 1.7\tim{-7} erg \cms\ s$^{-1}$, and if the source
was only 10\%\ coded by the BAT mask (56\deg\ off axis) the limit is 9.0\tim{-7} erg \cms\ s$^{-1}$.

The initial XRT observations (i.e. not including the observations of PS15dqa or PS15dpn)
covered 8.5 square degrees and enclosed 0.9\%\ of the probability in both the original and revised
sky maps\footnote{This rises to 12\%\ after galaxy convolution -- which was performed as we did not know at that time that
the source was a BBH.}. 55 sources were detected in these observations, however 39 of these were artefacts of an area of extended emission
(all but 2 of which were correctly flagged as such by the automated system and in the
counterpart notices, the final two were removed by visual inspection). Details of the 16 genuine sources, none of which is believed to be the counterpart,
are given in Table~\ref{tab:res_151226}; eight of these were rank 3 sources (uncatalogued, but below previous catalogue detection limits), and eight rank 4 (catalogued sources
at fluxes consistent with their catalogued values).

No uncatalogued sources were found in the UVOT data.

\subsubsection{Late observations}

On 2016 January 13 we performed a new set of observations of the error region of GW151226.
This consisted of 201 short ($\til60$ s) exposures, and was primarily performed as part of commissioning the ability to
rapidly tile GW error regions with \swift. These observations precede those reported in 
Section~\ref{sec:obs150914} (i.e. the 2016 April observations of GW150914),
and the test was only allowed to run for a few hours. This test revealed a bug in our software
affecting low resolution GW localisations ({\sc healpix}\footnote{http://healpix.sourceforge.net}-format maps with {\sc nside}$<512$), as a result
of which the 201 fields selected did not lie within the GW error region (this bug is now fixed).

As with the 2016 April 24-hour test (Section~\ref{sec:obs150914}), the only X-ray sources found were two
rank 4 sources. Fortuitously, these both lay in an area of the sky previously observed by \swift-XRT,
and the two sources were in the 1SXPS catalogue \citep{Evans14}, which allows us to compare their fluxes
with no spectral assumptions. The sources were 1SXPS J090436.8+553600 (catalogued XRT count-rate: $0.168\pm0.004$ s$^{-1}$, 
rate in the GW observations: $0.15\pm0.06$ s$^{-1}$) and 1SXPS J101504.1+492559 (catalogued at $1.300\pm0.008$ s$^{-1}$, 
observed at $2.0\pm0.4$ s$^{-1}$, i.e. both were consistent with the observed rate at the 1.5 $\sigma$ level).

\section{Optimisations for future GW runs}
\label{sec:future}

The second aLIGO observing run (`O2') is expected to take place in the second half of 2016, with
Advanced VIRGO (AdVIRGO; \citealt{Acernese15}) also anticipated to be collecting data during the latter part of this run
(the anticipated timeline for the aLIGO/AdVIRGO commissioning is given by \citealt{Abbott16f}).
As noted earlier, 
a new observing mode for \swift\ has now been commissioned, so it will be able to cover \til50 sq deg per day,
representing a significant improvement over the O1 response.

The core approach to \swift\ observations during O2 is expected to be as recommended by \cite{Evans16}: combining
the GW error region with an appropriate galaxy catalogue, and performing 60 s\footnote{\cite{Evans16} suggested
50-s exposures, but for technical reasons we cannot have observations shorter than 60 s.} observations
of as many of the most probable fields as possible, as soon as possible, for the first 48 hours
(when afterglow emission from an on-axis sGRB will be brightest). Thereafter we will re-observe these fields
for longer (500 s) exposures, as \cite{Evans16} argued that more than 48 hours after the GW trigger the population
of detectable sGRBs will be dominated by off-axis objects (which require longer observations to
detect). However, this broad plan hides a key detail: what galaxy catalogue should we use,
and indeed, how should we use it?

\subsection{Selecting a galaxy catalogue}
\label{sec:galcat}

For O1 we used the GWGC, since this extends to 100 Mpc, which the predictions
of \cite{Aasi13} and the simulations of \cite{Singer14} suggested was an appropriate
horizon for binary neutron star mergers detectable by aLIGO during O1. These same authors
predict the horizon distance will be higher (up to \til250 Mpc) during O2. The two
GW sources detected so far were both at much larger distances, \til500 Mpc.
As discussed in Section~\ref{sec:intro}, while these sources are BBH mergers which were
not believed to be strong EM emitters, the possible detection of a sGRB
by \emph{Fermi} coincident with GW150914 renders this uncertain, and it would be preferable
to be able to observe the error regions from such triggers. If we still wish to reduce
the sky area searched by using galaxy catalogues, we therefore need a catalogue
with a reasonable degree of completeness out to at  least 500 Mpc. However, 
when extending to such a distance the number of galaxies becomes so large that the benefits of targetting galaxies are diminished, therefore
some means of selecting which galaxies to target preferentially is needed.

One method was proposed by \cite{Gehrels16}, who noted that by selecting only
the brightest galaxies (those that produce 50\%\ of galactic light) the probability
of selecting the GW host galaxy is immediately reduced by 50\%, but the number
of galaxies one has to search is reduced by more than 50\%\ (around 68\%\ according to our analysis).
Our approach is similar to this. We do not reject the fainter galaxies,
but our galaxy map is weighted by the luminosity of the galaxy, and each possible XRT
field of view (over the whole sky) is assigned a probability:

\begin{equation}
\P\propto \sum_i{\frac{L_i}{L_{\rm tot}} \P_{\rm GW}}
\label{eq:origp}
\end{equation}

\noindent where $\P_{\rm GW}$ is the probability that the GW event lies within the XRT field
of view according to the GW skymaps, $L_{\rm tot}$ is the total luminosity contained in the
galaxy catalogue, and the summation is over all galaxies within the XRT field of view.
\swift\ pointings are performed approximately in decreasing order of probability (modulo observing constraints and
some optimisations to keep the time observing to time slewing ratio high), and the number of fields observed
is not based on the probability enclosed, but is limited by the amount of observing time committed to the follow up.

This process could be further optimised if the distance to the GW event is available promptly, so that only galaxies
at an appropriate distance are selected. \cite{Singer16} showed that 3D skymaps could be rapidly produced 
by aLIGO during O2. For these, each pixel in the skymap would contain not just a probability, but also the
parameters for how that probability is distributed in distance. This would allow equation (1) to be modified
to include the distance to the galaxy (which may itself be a probability distribution) and the distance dependence
of $P_{\rm GW}$, as we demonstrate shortly.

\begin{figure}
\begin{center}
\includegraphics[width=8.1cm,angle=0]{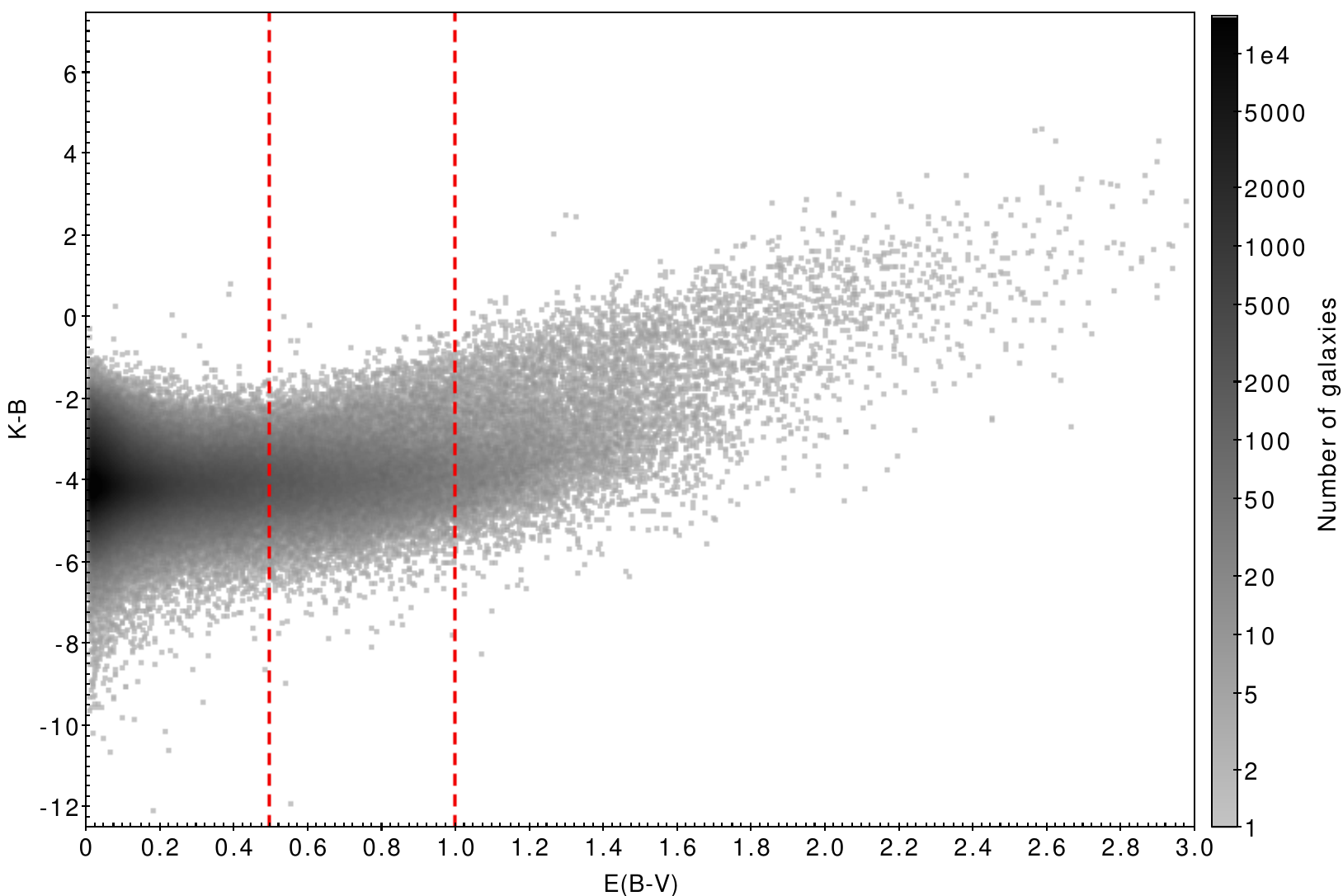}
\includegraphics[width=8.1cm,angle=0]{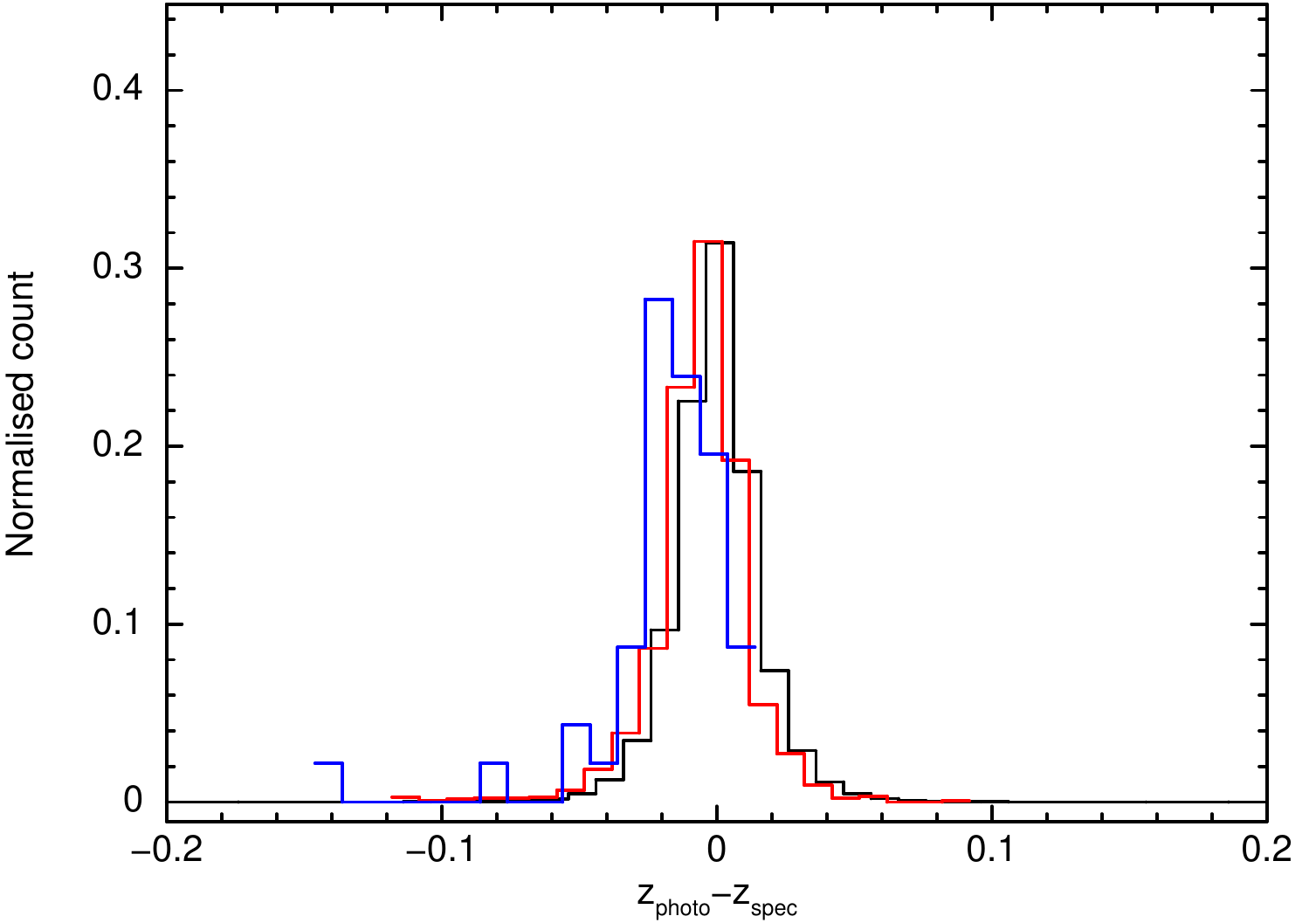}
\end{center}
\caption{The effect of extinction on the colour and redshift accuracy of the 2MPZ V1.1 catalogue.
\emph{Top}: The extinction-corrected $K-B$ colour as a function of extinction. For values
of $E(B-V)\ge0.5$ a clear trend starts to emerge, indicating a selection effect due to reddening.
The contours indicate the density of the points, and the red lines mark $E(B-V)=0.5$ and 1.
\emph{Bottom}: Histograms of $z_{\rm photo}-z{\rm spec}$ for objects with spectroscopic redshift measurements,
for $E(B-V)<0.5$ (black), $0.5\le E(B-V)<1$ (red) and $E(B-V)\ge1$ (blue). The red histogram shows
a small systematic shift, such that the mean photometric redshift is 0.004 too low compared to spectroscopic
values; the widths are comparible. At high extinction, the photometric redshift calibration becomes very poor.}
\label{fig:2MPZ}
\end{figure}

\begin{figure*}
\begin{center}
\includegraphics[width=8.5cm,angle=0]{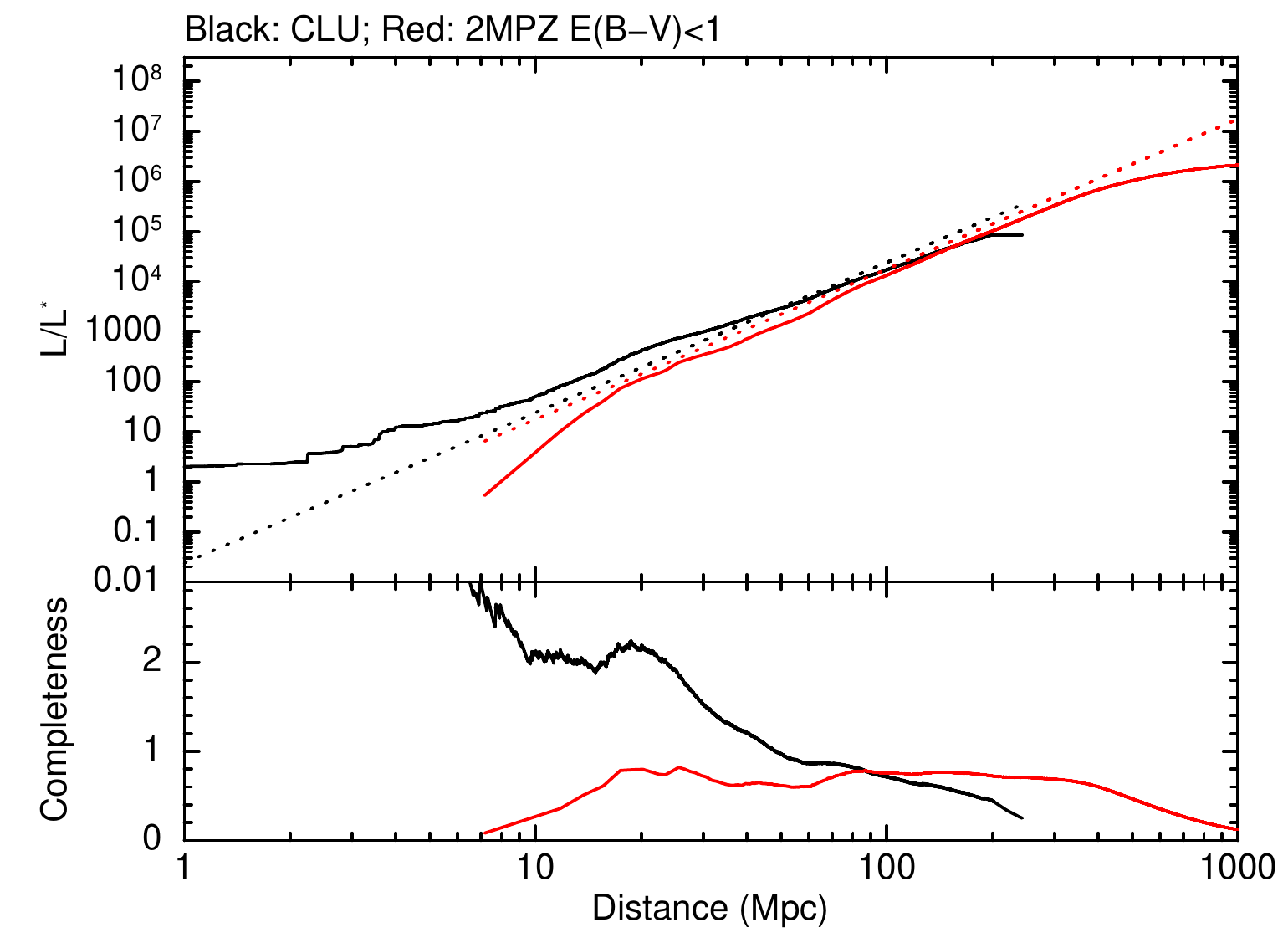}
\includegraphics[width=8.5cm,angle=0]{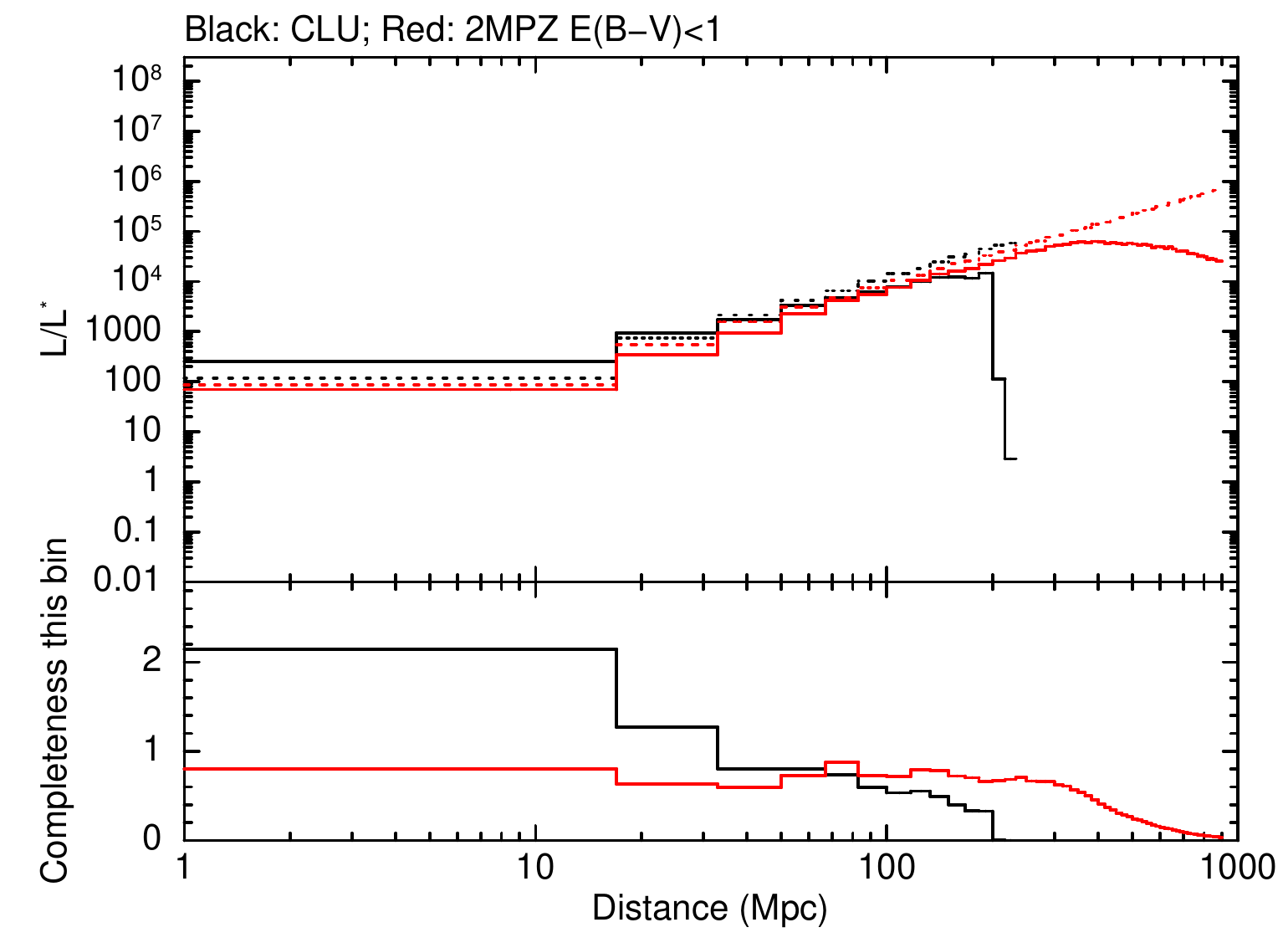}
\end{center}
\caption{The completeness of the CLU (black) and 2MPZ V1.1 (red). In each plot the upper pane shows the
theoretical total $L/L^*$ predicted by a Schechter function (dotted line) as a function of distance, and the observed value
from the catalogues (solid line); the lower pane shows the ratio of the theoretical to observed,
which we interpret as the completeness of the catalogue. 
The CLU data and theoretical values
relate to $B$-band magnitudes. The 2MPZ data use $K$-band magnitudes. For 2MPZ, only sources
in regions with $E(B-V)<1$ are included; the magnitudes
have been corrected from isophotal to total values; see text for details.
\emph{Left:} Luminosity is integrated out to the distance on the x-axis, hence completeness refers to how
complete the catalogue is out to distance D.
\emph{Right:} Luminosity is calculated in distance bins, hence completeness refers to how complete the catalogue
is at distance D.}
\label{fig:complete}
\end{figure*}

First, we must select an appropriate galaxy catalogue. Ideally this will be highly complete out to at least 500 Mpc,
have uniform sky coverage, and reliable luminosity (in a single band) and distance measurements for every galaxy.

\cite{Gehrels16} introduced the `Census of the Local Universe' (CLU) catalogue, and \cite{Evans16} suggested
the 2MASS Photometric Redshift catalogue [2MPZ; \cite{Bilicki14} and \cite{Antolini16} also suggested using 2MPZ].  The CLU is a meta catalogue created from 
several existing catalogues (Kasliwal, in preparation), whereas 2MPZ is based on a cross-correlation of the
2MASS extended source catalogue with the \emph{WISE} and \emph{SuperCOSMOS} all-sky catalogues.
Following \cite{Gehrels16} and earlier works
(e.g. \cite{White11}) we estimate the completeness of the catalogues by comparing the integrated luminosity
observed out to a given distance, with that predicted by a Schechter function \citep{Schechter76}. Using the terminology
of \cite{Gehrels16} we can define this function in terms of $x=L/L^*$ and the
integrated luminosity per unit volume is therefore given by:

\begin{equation}
L dV = \int_0^\infty \phi^* L^* x^{\alpha+1} e^{-x} dx
\label{eq:schech1}
\end{equation}

\noindent where $L^*$, $\phi^*$ and $\alpha$ are measured from observations. For the $B$-band data used in the
CLU\footnote{The CLU is a meta catalogue, so for component catalogues where $B$ is not available, a pseudo $B$ magnitude
is inferred and supplied.}, \cite{Gehrels16} give $M_B^*=-19.7 + 5\log h$, $\alpha_B=-1.07$ and $\phi_B^*=0.016 h^3$ ($M^*$ is 
the absolute magnitude of a galaxy with luminosity $L^*$); for 2MPZ we use the $K$-band magnitudes (as the catalogue is IR-selected) and the parameters from \mbox{\cite{Kochanek01}}:
$M_K^*=-23.39 + 5\log h$, $\alpha_B=-1.09$ and $\phi_B^*=0.0116 h^3$. We assumed $h=0.7(=H_0/100)$. To avoid questions
of photometric zeropoints, rather than comparing the observed luminosity with that predicted by the Schechter function,
we compare $x=L/L^*$, thus the zeropoints cancel out. From equation~\ref{eq:schech1}, the theoretical value for $x$
within volume $V$ is 

\begin{equation}
x = \int_0^\infty \phi^* x^{\alpha+1} e^{-x} dx V = \phi^*\Gamma(\alpha+2, 0) V
\label{eq:schech2}
\end{equation}

\noindent where $\Gamma$ is the incomplete gamma function.

The 2MPZ catalogue contains not the total infra-red magnitude of each galaxy, but 
instead those measured out to the 20mag/sq arcsec
isophote (for the $K$ band, labelled as `k\_m\_k20fe' in the 2MASS
database). Such magnitudes will systematically miss some flux and
need to be corrected for the total light. We follow \cite{Bilicki11}
and use the mean correction advocated by Kochanek et al. (2001),
subtracting 0.2 mag from the $K$-band magnitudes of every entry
in 2MPZ. We used a pre-release version 1.1
of 2MPZ\footnote{Now publicly available for download from the Wide
Field Astronomy Unit at the Institute for Astronomy, Edinburgh: http://surveys.roe.ac.uk/ssa/TWOMPZ}.
which contains \til6,000  extra galaxies compared to 2MPZ (added after the correction of \emph{WISE} instrumental 
artefacts), and the version we used contains more data that the public one
as it had no cuts made for Galactic extinction or stellar density.
\cite{Bilicki14} noted that such cuts are important to preserve uniformity. At high extinction the dust maps of 
\cite{Schlegel98} may saturate (i.e.\ become inaccurate); more significantly, at high extinction the intrinsically fainter
galaxies become undetectable, and this is a function of wavelength, so intrinsically redder galaxies tend to
be retained while bluer ones are lost, biasing the sample. In areas of high stellar density, where
galaxies and stars may be blended, the colours can also become unreliable.
For these reasons the publicly released catalogues (both v1 and v1.1) do 
not include sources for which $E(B-V)>1.5$ mag or log(stellar density, sources per square degree)$>$4.0;
no such filters were applied to the dataset we began with (providing an extra 6,700 
sources compared to the public dataset):
we wish for a sample that is as homogeneous as possible, 
yet also accurate and complete, therefore we explored what cuts were necessary to achieve this.
Fig.~\ref{fig:2MPZ} shows the $K-B$ colour (both values are corrected for Galactic extinction)\footnote{In 2MPZ the magnitudes
are corrected for Galactic extinction, but not for cosmological effects (i.e. the $k$-correction). The 2MASS and \emph{WISE}
magnitudes are in the Vega system, 
whereas the \emph{WISE} and \emph{SuperCOSMOS} magnitudes are AB-like magnitudes (Peacock et al.\ in prep).}
as a function of extinction, $E(B-V)$. A clear bias begins to 
emerge when the extinction exceeds 0.5 mag, demonstrating that the 
extinction correction, and therefore colours and hence photometric 
redshifts ($z_{\rm photo}$), are unreliable in this regime. The bottom 
panel of Fig.~\ref{fig:2MPZ} shows the difference between the photometric
redshift and spectroscopic redshift ($z_{\rm spec}$) for those objects with both,
for three samples: $E(B-V)<0.5$, 
$0.5\le E(B-V)<1$ and $E(B-V)\ge1$. The $0.5\le E(B-V)<1$ 
sample is slightly off-centre, suggesting that at these extinctions, 
$z_{\rm photo}$ is systematically underestimated by 0.004, however this shift is small compared
to the uncertainty (discussed below) and can be ignored. 
There are only 46 objects in 2MPZ with a spectroscopic redshift and  $E(B-V)\ge1$ so we have not
subdivided the data further for Fig.~\ref{fig:2MPZ}, however the effect
of extinction on the distribution of photometric redshift is similar to that in
Fig.~\ref{fig:2MPZ}; at higher extinctions, the distribution is biased towards ever lower
photometric redshifts. While the number of galaxies at high extinctions is small
compared to the overall catalogue, this clustering means that the inclusion of 
high-extinction objects significantly distorts the measurements of how complete
the catalogue is.
We therefore applied 
a cut of  $E(B-V)<1$ to 2MPZ, resulting in 
9,638 (1.0\%) of the sources being discarded; all future discussion of 2MPZ in this work
refers to this sample. We also investigated the effect
that stellar density has on the accuracy of the catalogue. As with extinction,
a clear bias in colour is visible at high stellar densities, however
the extinction filtering just described removes all the sources
with colours affected by stellar density, so an independent density filter is not needed.

In Fig.~\ref{fig:complete} we show the completeness of CLU and
2MPZ. At $D<40$Mpc, CLU is over-complete (i.e.\ contains more than the
expected luminosity): this was also true of GWGC which
\cite{White11} attributed to the effect of the Virgo cluster.  2MPZ does not show
this overcompleteness, this is likely the result both of the (comparatively) low sensitivity of
2MASS to low-surface-brightness galaxies (which dominate the nearby sample), and the inaccuracy 
of the photometric redshift: we return to the latter point below.
Beyond \til60 Mpc the 2MPZ survey is significantly more complete
than CLU. 2MPZ also has the advantage of being more uniform across the sky than CLU; compare
fig.~13 of \mbox{\cite{Bilicki14}} with fig.~1 of \mbox{\cite{Gehrels16}}. In particular, completeness of CLU will depend on sky
position, as this dataset is constructed from spectroscopic surveys,
which at present do not cover the full sky beyond $\til$130 Mpc [$z=0.03$:
completeness of the 2MASS Redshift Survey, \cite{Huchra12}]. Such
a limitation does not apply to photometric all-sky surveys, such as
2MASS or \emph{WISE}, which are only limited by their respective flux limits
and Galactic plane nuisances.

These considerations suggest that 2MPZ represents the better catalogue to use, although if the GW information
shows that the object is $<60$ Mpc away it may be better to instead use CLU. 

\begin{figure}
\begin{center}
\includegraphics[width=8.1cm,angle=0]{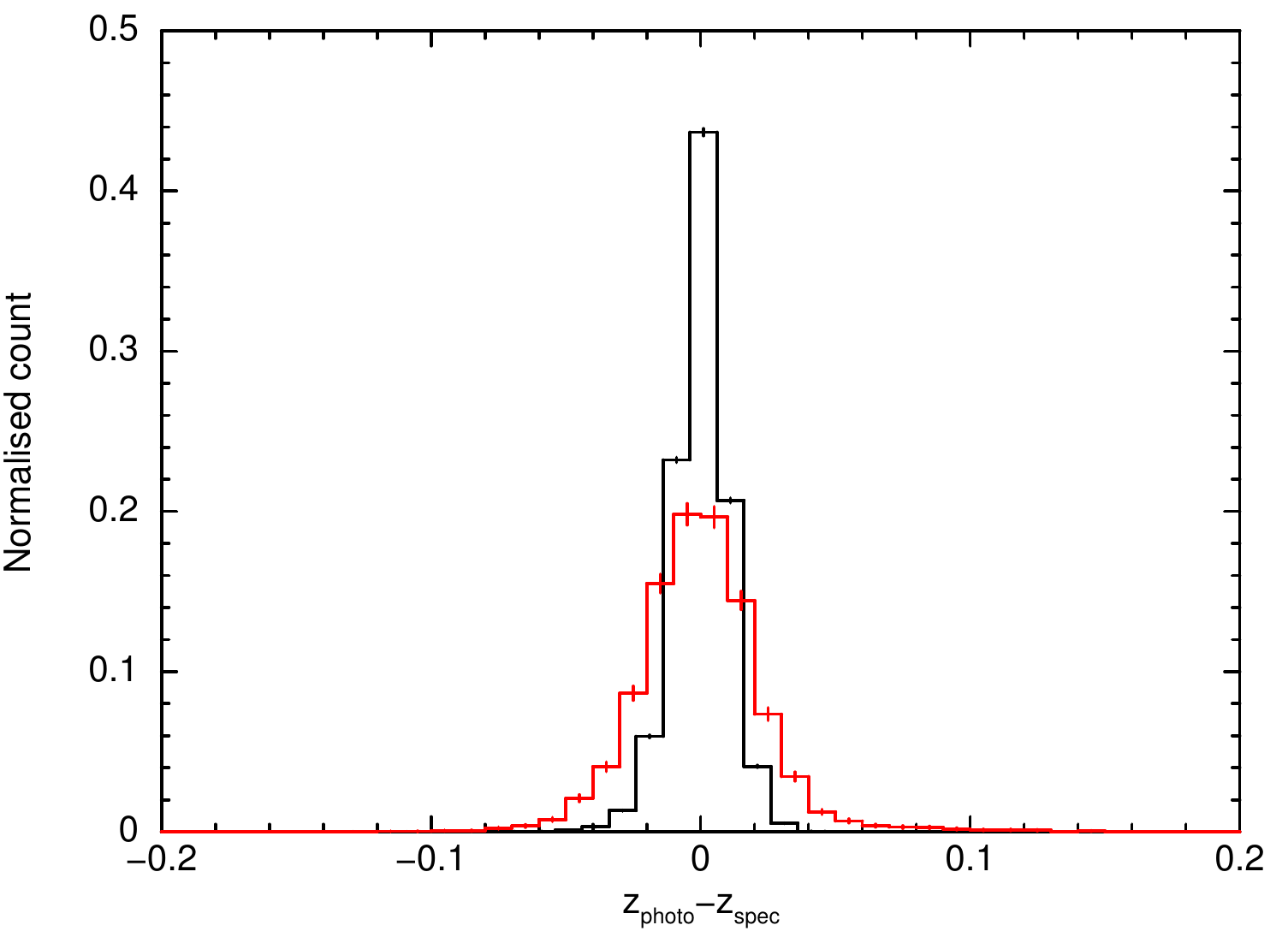}
\includegraphics[width=8.1cm,angle=0]{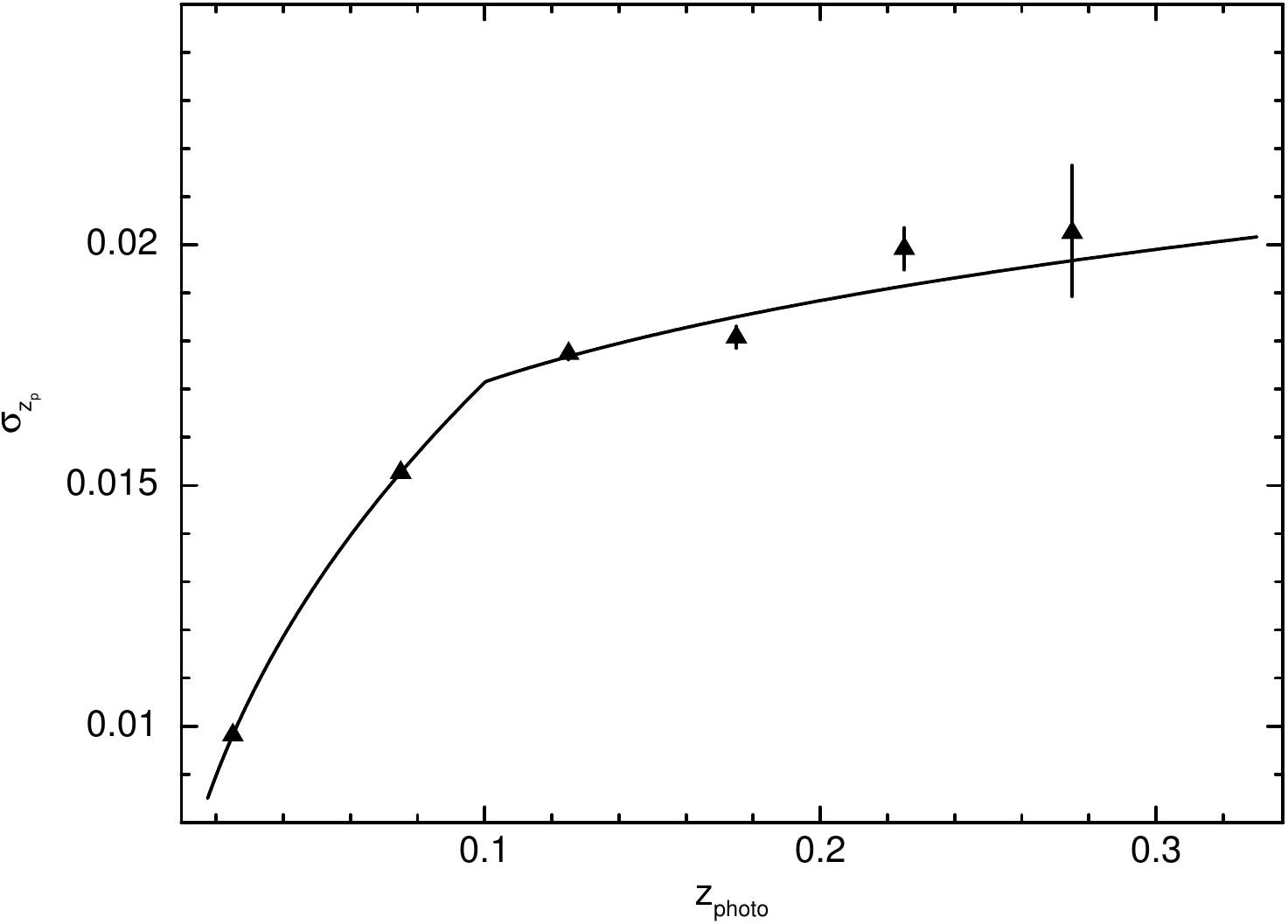}
\end{center}
\caption{The accuracy of the photometric redshifts in 2MPZ. \emph{Top:} Histograms
of $z_{\rm photo}-z_{\rm spec}$ for galaxies with $z_{\rm photo}<0.05$ (black) and
$0.2\le z_{\rm photo} < 0.25$ (red). \emph{Bottom:} The variation of the Gaussian $\sigma$
of the histograms as a function of photometric redshift; the model is given in equation~\ref{eq:sigma}.}
\label{fig:photoz}
\end{figure}

However, the accuracy of 
the redshift information must also be considered. While CLU employs only spectroscopic redshift measurements,
roughly 2/3 of the sources in 2MPZ have only photometric redshifts, which will have a larger uncertainty.
To determine the accuracy of the photometric redshifts, we selected from 2MPZ only those objects with 
$E(B-V)<1$ and with both spectroscopic and photometric redshifts (the latter had the systematic correction 
above applied for $E(B-V)\ge0.5$). We then created histograms of $z_{\rm photo}-z_{\rm spec}$
in several $z_{\rm photo}$ bins; two examples are shown in  Fig.~\ref{fig:photoz}. These are 
approximately Gaussian, but the width ($\sigma_{z_p}$) varies with redshift. We fit this
variation (Fig.~\ref{fig:photoz}, lower panel) with a broken powerlaw:

\begin{equation}
\sigma_{z_p} = \left\{
\begin{array}{ll}
0.043\ z_{\rm photo}^{0.402} & z_{\rm photo} < 0.10 \\
0.023\ z_{\rm photo}^{0.14} & z_{\rm photo} \ge 0.10 \\
\end{array}
\right.
\label{eq:sigma}
\end{equation}

\noindent as shown in the lower panel of Fig.~\ref{fig:photoz}. This uncertainty must be taken into account
when comparing the distance of a galaxy with the distance inferred from the GW data, as will be
described in Section~\ref{sec:galdist}. This relatively large uncertainty in photometric redshift will also
distort the completeness curve slightly. The probability of a galaxy with
true redshift $z_1$ being assigned $z_{\rm photo}=z_1+\dz=z_2$ is slightly lower than the inverse:
the probability of a galaxy with a true redshift $z_2$ being assigned $z_{\rm photo}=z_2-\dz=z_1$, because
there is more volume (and so more galaxies) at $z_2$ than at $z_1$\footnote{This is analogous to
the Malmquist bias present for redshift-independent distance indicators and peculiar velocities derived from them.}.
This is clearly a bigger effect at lower redshift,
where $\Delta z/z$ is higher. Similarly towards the limit of the catalogue's redshift range (i.e. $z\to0$ and $z\to z_{\rm lim}$)
the completeness will be underestimated because some galaxies with true redshifts inside the catalogue's redshift range will
receive photometric values outside of it, but there are no (or few) galaxies outside of the limit to `compensate'.
The overall effect of this is that the completeness shown in Fig.~\ref{fig:complete} is underestimated
until the catalogue limit ($z\til0.3$ corresponding to $D\til 1.6$ Gpc) is approached, i.e.\ 2MPZ is more complete
within the distance range we are interested in than implied from Fig.~\ref{fig:complete}, which strengthens the argument that 2MPZ is 
the better choice of the two catalogues we have studied. Looking further in the future as the horizon distance
of aLIGO/AVIRGO rises further, we may wish to consider the forthcoming WISExSuperCOSMOS photometric
redshift catalogue (Bilicki et al. 2016 in press), which 
covers less of sky than 2MPZ (70\%), but reaches much deeper in
redshift (median $z\til0.2, D\til850$ Mpc), however we defer study on the cost/benefit of this to a future work.

\subsection{Using the distance and completeness}
\label{sec:galdist}

In our GW response to date, we have used galaxy catalogues in a simplistic way: we ignored the (in)completeness
of the catalogue (i.e. regions of the sky without known galaxies were given zero probability of hosting the GW event)
and did not weight galaxies by their distance compared to the expected distance to the GW event -- the
latter was not possible due to the lack of GW distance estimate available
at trigger-time. In O2 and beyond, the horizon distance is such that the incompleteness of galaxy catalogues
is significant (Fig.~\ref{fig:complete}), and the aLIGO/VIRGO teams are likely to produce rapid
distance estimates \citep{Singer16}; therefore, we describe now a new method of galaxy convolution to produce
higher-fidelity skymaps than we have used to date.

Considering first completeness: if the distance $D$ to the GW source is known 
perfectly, then we can estimate the completeness of the galaxy 
catalogue at this distance from the lower panel of Fig.~\ref{fig:complete}, we will call this $C$. The 
probability that the GW event occurred in a known galaxy is thus $C$, and
the probability that it occurred in an unknown one is $1-C$. 
\cite{Singer16} demonstrated that the distance $D$ deduced from the GW data is
a function of direction on the sky. The GW error regions are distributed as
{\sc healpix}-format skymaps, and each pixel in this map in the \cite{Singer16} approach
has its own $D$ distribution and hence $C$ value, which
we calculate thus\footnote{Note that we are implicitly assuming that the completeness
of the catalogue is not a function of direction. This is a reasonable assumption
for 2MPZ, but for current spectroscopic surveys the non-uniformity on the sky
would need to be factored into $C_p$.}:

\begin{equation}
\label{eq:cpix}
C_p = \frac{\int  P_p(D) C(D) dD }{\int P_p(D) dD}
\end{equation}

\noindent where $P_p(D)$ is the probability distribution of the distance, defined for the pixel $p$. 
Therefore, for each given pixel in the skymap, the probability of the GW event occurring in
an uncatalogued galaxy within that pixel is:

\begin{equation}
\P_{{\rm nogal},p}=\P_{{\rm GW},p} \left( 1-C_p \right)
\label{eq:pnogal}
\end{equation}

\noindent where $\P_{{\rm GW},p}$ is the probability in the original skymap from the aLIGO/AVIRGO team for pixel $p$.

For pixels containing galaxies, an extra factor $\P_{{\rm gal},p}$ must be included.
Previously (paper I) we defined this as in equation~\ref{eq:origp}: the GW probability multiplied by 
the ratio of galaxy luminosity in the pixel to the total catalogued galaxy luminosity. This now
needs to refer only to the luminosity within the distance indicated by the GW dataset, and
needs a correction for completeness. We therefore redefine the probability of the GW event
occurring within a known galaxy in pixel $p$ thus\footnote{The normalisation of this equation
was incorrect in the paper as originally published. This has been corrected in an erratum, this 
version contains the corrected equations.}:

\begin{equation}
\P_{{\rm gal},p}=\P_{{\rm GW},p} C_p N\sum_g \left( \P(g|P_p[D]) \frac{L_g}{L_{\rm tot}} \right)  
\label{eq:pgal}
\end{equation}

\noindent The summation is over all galaxies $g$ in pixel $p$. $L_g$ is the luminosity of galaxy $g$
divided by the number of pixels it covers. $N$ is a normalisation needed to ensure that $\sum_p \P_{{\rm gal},p}=C$
(if $C$ is the catalogue completeness averaged over all pixels). This is given by:
  
\begin{equation}
N=\frac{ \sum_p \P_{{\rm GW},p} C_p }  {\sum_p \left( C_p \P_{{\rm GW},p}  \sum_g{\left( \P(g|P_p[D]) \frac{L_g}{L_{\rm tot}}\right)  }\right)}
\label{eq:norm}
\end{equation}

\noindent $L_{\rm tot}$ in equations (\ref{eq:pgal})--(\ref{eq:norm}) is the total catalogued galaxy luminosity within the GW volume, so  $\sum_g\frac{L_g}{L_{\rm tot}}$
gives the ratio of the actual luminosity in pixel $p$ compared to the total distance-weighted luminosity of all galaxies in the GW error region,
i.e. the relative probability of the galaxies in this pixel containing a merger event compared to those in any other pixel.
$L_{\rm tot}$ is given by:

\begin{equation}
L_{\rm tot} = \sum_p \sum_g \left[ \P(g|P_p[D]) L_g \right]
\label{eq:ltot}
\end{equation}

\noindent where $\P(g|P(D)_p)$ is the probability that the galaxy $g$ is at the correct
distance to host the GW event. This is simply:

\begin{equation}
\P(g|P_p[D]) = \int P_p(D) P_g(D) dD
\label{eq:pgal_d}
\end{equation}

\noindent where $P(D)_p$ is the probability as a function of distance for pixel P, determined from the
GW data. For the low-latency analysis this is a Gaussian multiplied by distance squared \citep{Singer16}.
$P_g(D)$ is the probability distribution of the distance $D$ of galaxy $g$. 2MPZ does not contain uncertainties
on the photometric redshift measurements, therefore we need to decide on the form of $P_g(D)$.
For galaxies in 2MPZ with spectroscopic redshift we assume that the dominant source of
error is the peculiar velocity of the galaxy, and we take 500 km s$^{-1}$ as representative
of this. This corresponds to a distance error of $500/H_0 = 7.4$ Mpc (assuming $H_0=70$ km s$^{-1}$ Mpc$^{-1}$), so for
galaxies with spectroscopic redshift we treat $P_g(D)$ as a Gaussian with $\sigma=7.4$ Mpc.
For photometric redshifts we use the prescription given in Section~\ref{sec:galcat}
(the peculiar velocity correction is insignificant compared
to this and can be ignored).

Having now calculated the probability of the GW event occurring in an unknown galaxy, or in any specific known galaxy, the
probability that the GW event is in pixel $p$ of the skymap is simply:

\begin{equation}
\P_p=\P_{{\rm nogal},p}+\P_{{\rm gal},p}
\label{eq:finalProb}
\end{equation}

\noindent Finally, the map must be renormalised such that it sums to unity\footnote{Renormalisation
is not essential for planning observations, since it is the relative probability in each pixel that matters;
however in order to calculate the probability that one has observed the true GW location, the map 
must be normalised to 1.}. The result of this is a modified probability map on the sky
which accounts for both the GW localisation and our prior knowledge of the structure of the local 
universe. This can then be used in a manner similar to that proposed by \cite{Gehrels16}, i.e.\ by selecting
fields in (descending) probability order until some threshold probability has been selected (50\%\ in the method
of \citealt{Gehrels16}). Alternatively, as suggested in Section~\ref{sec:future}, we can observe as many fields as possible
in a given time interval, but again observing in order of priority.

Based on local structure in the universe it could be argued
that the unknown galaxies are not homogeneously distributed on the sky, but are instead more likely near known 
galaxies. In this case, more exotic definitions of $P_{{\rm nogal},p}$ could be created to account for the distance to nearby
galaxies. For the present, we will limit ourselves to the simple prescription above.
Similarly since the values of binary inclination and distance determined from the
GW are degenerate, the probability distribution of binary inclination for each pixel could in principle be produced
and then, from a template library of GRB light curves for different inclinations, one could determine
the probability of detecting a GRB from a given pixel as a function of time. While this is under investigation it
is not likely to be possible on the timescale of O2, and is beyond the scope of this paper.

\subsection{Which luminosity to use}
\label{sec:band}

The above calculations weigh each galaxy by its luminosity. However, which band one uses is also pertinent\footnote{Technically one should consider the rest-frame band
rather than the observer frame, however since we are considering only the relatively nearby universe, we neglect this issue.}.
The $K$-band provides a reasonable proxy for stellar mass in the galaxy, which we may take as being a proxy for the
number of binary neutron-star systems in the galaxy and hence the probability of hosting a merger of such a system.
However, recent observations \citep{Davanzo09,Fong13} suggest that short GRBs are more common in late-type galaxies (suggesting
that the probability of a compact binary coalescing is influenced by recent star formation), which suggest that it is more
appropriate to weight galaxies by their $B$ band luminosity. 2MPZ contains both infrarad magnitudes (from 2MASS and \emph{WISE}) 
and the optical $R$ and $B$ magnitudes (from SuperCOSMOS), which gives us the flexibility to select which band we wish
to use, if the theoretical (or observation) priors change.

To investigate this, and the impact of the galaxy convolution, we performed a series of simulations. We started with the
GW simulations of \cite{Singer16}\footnote{https://dcc.ligo.org/public/0119/P1500071/005/index.html},
which provide 3-D probability maps for 250 simulated binary neutron star mergers in the 2-detector configuration\footnote{Labelled as `O1' online.}.
These simulations assume that the mergers are simply distributed homogeneously in space, whereas we wish to seed them in galaxies.
Each GW simulation has the position and distance to the simulated event. We calculated the completeness $C$ of 2MPZ at this distance,
and then generated a random number $0\le\mathcal{R}<1$. If $\mathcal{R}\ge C$ the GW event was treated as occuring in
an uncatalogued galaxy, so the data needed no changes. Otherwise, a host galaxy for the event was selected at random from the 2MPZ
catalogue, with each galaxy having a probabilty of being the host, proportional to $L\mathcal{P}(D)$ (where $L$ is the galaxy
luminosity and $\mathcal{P}(D)$ is the probability that this galaxy is at the distance of the simulated merger).
Since the LIGO probability maps are strongly dependent on the geocentric direction to the merger, rather than rotate these
maps such that the GW events occured in galaxies, instead we rotated the galaxy catalogue such that the selected host was
at the position of the simulated merger. We then created a series of XRT fields tiled on the sky, arranged them in decreasing 
order of probability and determined which field contained the merger event.

We did this five times with different models. In the first instance, we performed no galaxy convolution at all, i.e. we simulated tiling the original GW error region.
In the other four simulations, we selected host galaxies based on either their $B$- or $K$-band luminosities, and then convolved
the GW map with 2MPZ using the $B$- or $K$-band luminosities. That is, we simulated the cases where our assumption
about which galaxies are more likely to host GW events are correct (the same band was used in selecting the host and convolving the GW region),
and when they are incorrect (one band was used to select the hosts, and the other used in convolution).

\begin{figure}
\begin{center}
\includegraphics[width=8.1cm,angle=0]{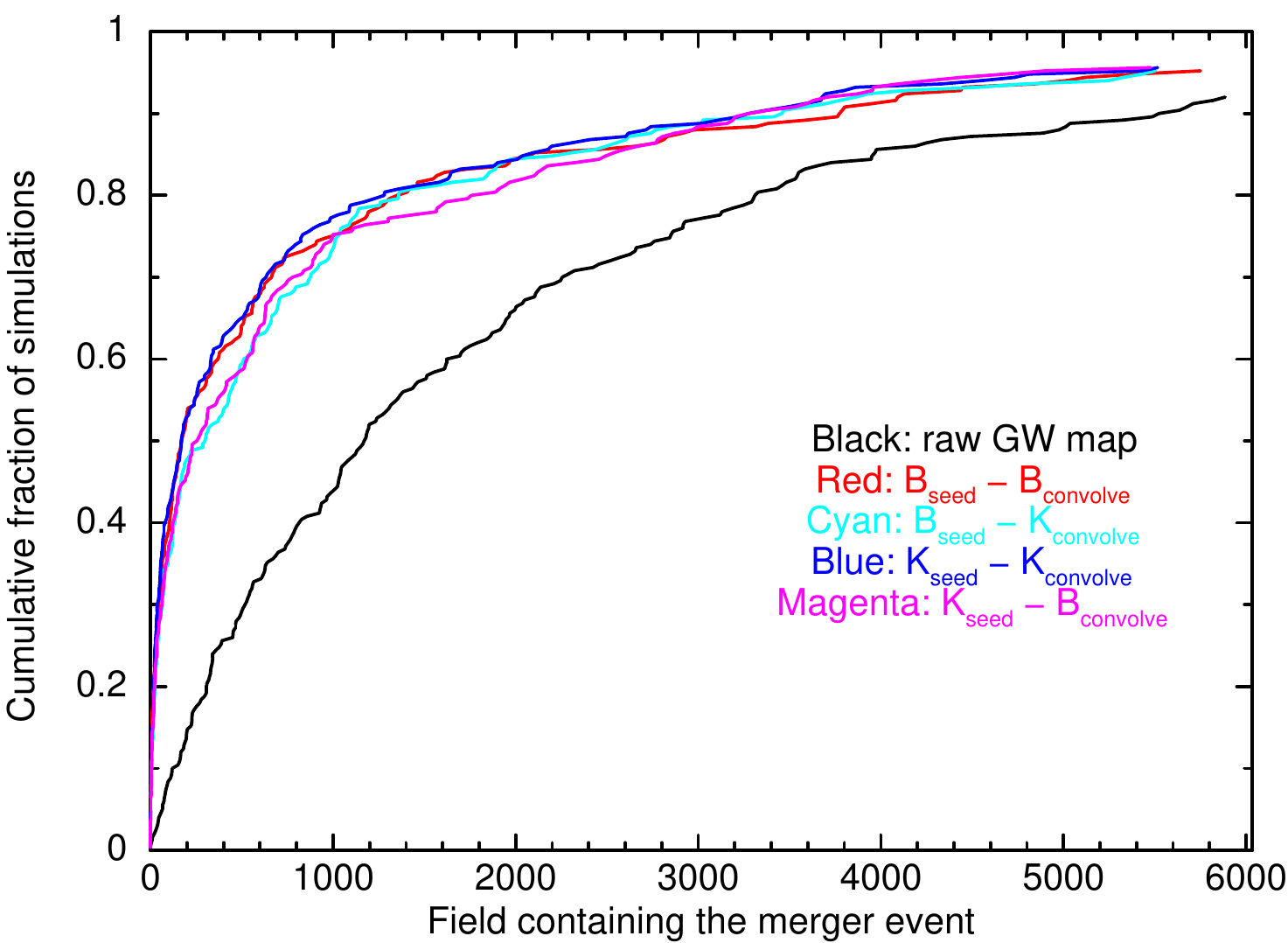}

\end{center}
\caption{The cumulative distribution of which XRT field (in probability order) contains the GW event,
from 250 simulated GW events (per colour), showing the benefit of galaxy convolution and the impact of
using an incorrect assumption as to which luminosity band the galaxies should be weighted by.
\emph{Black:} Simulations where the XRT fields are generated
based on the original GW map with no galaxy convolution. \emph{Red:} The GW events are seeded in hosts
weighted by $B$-band luminosity, and the GW error region is convolved with a $B$-band weighted galaxy catalogue.
\emph{Cyan:} Seeding is weighted by the $B$-band, but convolution uses the $K$-band. \emph{Blue}: Seeding and convolution
both use the $K$-band. \emph{Magenta:} Seeding is weighted by the $K$-band and convolution by the $B$-band.
}
\label{fig:simresults}
\end{figure}

In Fig.~\ref{fig:simresults} we show the results of this. Plotted is the cumulative distribution of which field contained the GRB
in the 250 simulated mergers, for the different simulations runs. This confirms that
the galaxy convolution significantly reduces the typical number of XRT fields we have to observe before we reach the correct
location. With no convolution, 50\%\ of the time at least 1,200 XRT fields are needed to reach the correct location. 
With convolution this falls to \til170 (a factor of \til7 decrease) if the same band is used
in the simulations and search, or \til 300 fields (a factor of \til4 improvement over the no-convolution approach)
if different bands are used. Therefore, while the choice of which band to use when convolving galaxies does have a
significant effect, choosing the wrong band is still much better than not using galaxy convolution.

\section{Conclusions}
\label{sec:conc}

\swift\ performed rapid-response follow up to all three GW triggers released to the EM
partners by the aLIGO team during the O1 operating run of aLIGO. No compelling X-ray, optical
or gamma-ray counterpart was found, however this is not surprising, since only a small fraction
of the GW error region was covered. Additionally, one of the GW triggers was spurious and the
other two are believed to be BBH mergers, which may not be expected to give rise to EM emission.
For the second trigger, we can place a limit on the hard X-ray emission of (4.3--90)\tim{-8}  erg \cms\ s$^{-1}$ (15--350 keV)
for a region enclosing 15\%\ of the GW probability.

In the future \swift\ will be able to observe a much larger fraction of the GW error region
as a new observing capability has been commissioned, which will enable  
large-scale, short-exposure tiling. Given both the increased horizon distance 
expected during O2, and the fact that both real GW events in O1 were at large
distances (\til500 Mpc), targeting galaxies in the GWGC, which is limited to 100 Mpc, is 
not a good approach. The 2MPZ catalogue, which uses a mixture of spectroscopic and photometric redshifts,
offers a better prospect (unless the GW localisation identifies
the object as being $<60$ Mpc from Earth), and we have shown how we can use the completeness
measurements for this catalogue, and the GW distance estimates expected to be rapidly available
in O2, to optimise the skymap produced by the aLIGO/AVIRGO teams. In the future, as new catalogues
become available (such as the WISExSuperCOSMOS, GLADE in prep/press) or when photometric redshifts
are added to Wise-2MASS complilation \citep{Kovacs15}, 
and  the sensitivity and localisation characteristics of aLIGO/AVIRGO improve it may be valuable
to reassess the benefits of galaxy targeting and choice of catalogue.

\section*{Acknowledgements}

We thank Andr\'as Kov\'acs for helpful discussion on galaxy catalogues. 
This work made use of data supplied by the UK Swift Science Data Centre 
at the University of Leicester. This publication 
makes use of data products from the Two Micron All Sky Survey, which is 
a joint project of the University of Massachusetts and the Infrared 
Processing and Analysis Center/California Institute of Technology, 
funded by the National Aeronautics and Space Administration and the 
National Science Foundation. This research has made use of the  XRT Data 
Analysis Software (XRTDAS) developed under the responsibility of the ASI 
Science Data Center (ASDC), Italy. This research has made use of the SIMBAD database,
operated at CDS, Strasbourg, France. The GW probability maps and our related
galaxy maps are in {\sc healpix} format \citep{Gorski05}.
PAE, JPO and KLP acknowledge UK Space Agency support. 
SC and GT acknowledge ASI for support (contract I/004/11/1).
MB is supported by the Netherlands Organization for Scientific Research, NWO, through grant number 614.001.451;
through FP7 grant number 279396 from the European Research Council; and by the Polish National Science
Centre under contract \#UMO-2012/07/D/ST9/02785.

\bibliographystyle{mnras} \bibliography{phil}

\label{lastpage}

\end{document}